\documentclass[fleqn,10pt]{wlscirep}
\usepackage[utf8]{inputenc}
\usepackage[T1]{fontenc}

\usepackage{amsmath,amsthm,amssymb,amsfonts}
\usepackage{comment}
\usepackage{graphicx}
\usepackage{pdfpages}
\usepackage{algorithm}
\usepackage{algpseudocode}
\usepackage{appendix}
\usepackage{siunitx}
\usepackage{booktabs}
\usepackage{multirow}
\usepackage{tabularx}
\usepackage{bbm}
\usepackage{subcaption}
\usepackage{cases}
\usepackage{mathtools}

\usepackage[utf8]{inputenc}
\usepackage[english]{babel}

\title{Learning Plasma Dynamics and Robust Rampdown Trajectories with Predict-First Experiments at TCV}
\author[1,2,*]{Allen M. Wang}
\author[3]{Alessandro Pau}
\author[1]{Cristina Rea}
\author[2]{Oswin So}
\author[2]{Charles Dawson}
\author[3]{Olivier Sauter}
\author[4]{Mark D. Boyer}
\author[3]{Anna Vu}
\author[3]{Cristian Galperti}
\author[2]{Chuchu Fan}
\author[3]{Antoine Merle}
\author[3]{Yoeri Poels}
\author[3]{Cristina Venturini}
\author[3]{Federico Felici}
\author[3]{Stefano Marchioni}
\author[5]{the TCV Team}

\affil[1]{Plasma Science and Fusion Center, Massachusetts Institute of Technology, Cambridge, MA 02410, USA}
\affil[2]{Laboratory for Information and Decision Systems, Massachusetts Institute of Technology, Cambridge, MA 02410, USA}
\affil[3]{Swiss Plasma Center, Ecole Polytechnique F\'ed\'erale de Lausanne, CH-1015 Lausanne, Switzerland}
\affil[4]{Commonwealth Fusion Systems, 117 Hospital Rd, Devens, MA 01434}
\affil[5]{See the author list of B. P. Duval et al 2024 Nucl. Fusion 64 112023 \cite{duval2024experimental}}
\affil[*]{Corresponding Author: allenw@mit.edu}
\begin{abstract}
The rampdown phase of a tokamak pulse is difficult to simulate and often exacerbates multiple plasma instabilities. To reduce the risk of disrupting operations, we leverage advances in Scientific Machine Learning (SciML) to combine physics with data-driven models, developing a neural state-space model (NSSM) that predicts plasma dynamics during Tokamak à Configuration Variable (TCV) rampdowns. The NSSM efficiently learns dynamics from a modest dataset of 311 pulses with only five pulses in a reactor-relevant high-performance regime. The NSSM is parallelized across uncertainties, and reinforcement learning (RL) is applied to design trajectories that avoid instability limits. High-performance experiments at TCV show statistically significant improvements in relevant metrics. A predict-first experiment, increasing plasma current by 20\% from baseline, demonstrates the NSSM's ability to make small extrapolations. The developed approach paves the way for designing tokamak controls with robustness to considerable uncertainty and demonstrates the relevance of SciML for fusion experiments.
\end{abstract}
\begin{document}
\flushbottom
\maketitle
\thispagestyle{empty}

\section*{Introduction}
\begin{figure}[!htbp]
    \centering
    \includegraphics[width=\linewidth]{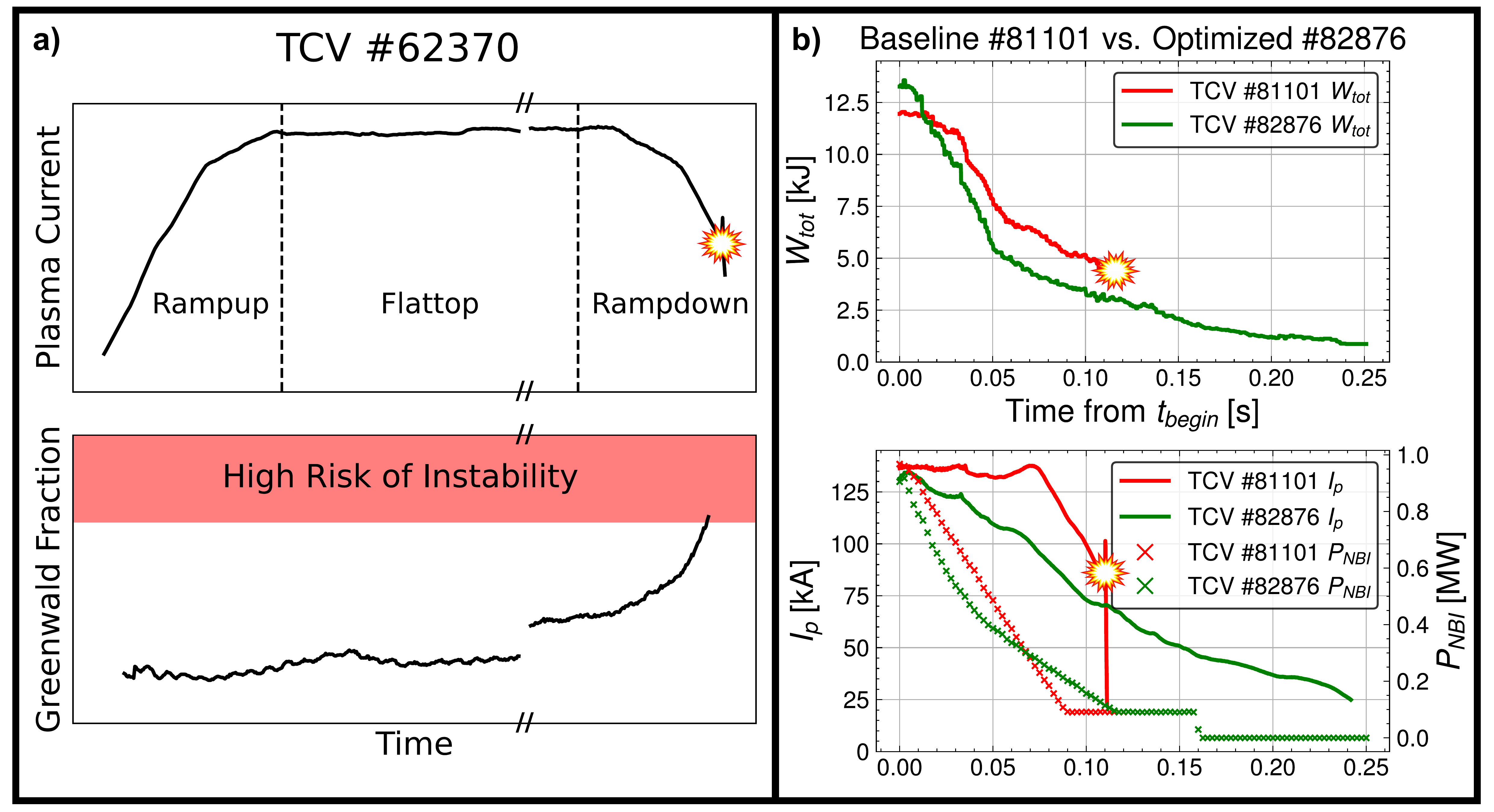}
    \caption{\textbf{Examples of plasma disruptions during rampdowns, and an example non-disruptive result obtained after deployment of the developed method}. \textbf{a)} Data from an illustrative TCV pulse (\#62370), showing the rampup, flattop, and rampdown phases, which are defined by the plasma current $I_p$. The bottom subplot shows how the rampdown pushes the plasma closer to an instability limit, in this case the Greenwald density limit, defined by the Greenwald fraction $f_{GW} = 1$. Note this limit is approximate due to the, at present, incomplete physics understanding of the density limit \cite{giacomin2022first, maris2024correlation}. Also note that the flattop phase is abbreviated here to more clearly highlight the rampup and rampdown. \textbf{b)} A comparison of the current, $I_p$,  stored energy, $W_{tot}$, and neutral beam injection power, $P_{NBI}$, trajectories for a baseline shot and an optimized shot, showing a faster, and non-disruptive, decrease in the plasma current and stored energy relative to the baseline.}
    \label{fig:intro-fig}
\end{figure}

Upcoming burning plasma tokamaks, such as SPARC \cite{creely2020overview} and ITER \cite{shimada2007overview}, require reliable plasma control to avoid operational delays and machine damage due to plasma disruptions, a challenge that will only increase for tokamak pilot plants \cite{maris2023impact} like ARC \cite{sorbom2015arc} and DEMO \cite{federici2014overview}. Given this risk becomes intolerable at high plasma current, $I_p$, and stored energy, $W_{tot}$, a key mitigation strategy is to de-energize the plasma by performing a rampdown of the plasma current, but doing so typically pushes the plasma closer to multiple instability boundaries \cite{van2023scenarioA, teplukhina2017simulation,mehta2024automated}. Figure \ref{fig:intro-fig} depicts the phases of a tokamak pulse, beginning with rampup of the plasma current to the steady-state flattop phase, and ending with a rampdown. Notably, Figure \ref{fig:intro-fig} also shows an example of a quantity correlated with plasma instability growing during the rampdown phase, a challenge which motivates the algorithmic design of safe rampdown trajectories. This challenge is especially acute in reactor relevant high performance plasmas, which operate close to instability boundaries to achieve the high normalized plasma density, typically quantified by the Greenwald fraction $f_{GW}$ \cite{greenwald1988new}, and normalized plasma pressure, $\beta_N$ \cite{troyon1984mhd}, important for economical energy production. The importance of designing robust rampdowns for reactor relevant fusion plasmas is highlighted by the recent record-breaking high performance campaign at the Joint European Torus (JET), for which most disruptions occurred during the termination phase \cite{garzotti2023development}. For the baseline scenario, a $\approx15\%$ increase of the plasma current, from 3MA to 3.5MA, increased the disruptivity considerably from $\approx20\%$ to $\approx50\%$ \cite{garzotti2023development}. This challenge motivates the development of tools that can rapidly adapt rampdown trajectories to manage disruptivity as fusion performance is increased.

Due to the stochasticity of plasma dynamics, hardware and control imperfections, and the possibility of off-normal-events (ONEs), it is important to design scenarios, trajectories, and controllers with robustness to distributional uncertainty in the dynamics of the plasma. The biggest barrier to designing for robustness in this context is the difficulty of simulating plasma dynamics during the highly transient rampdown phase, during which multiple physical quantities, many of which are not well-modeled with a principles-based approach, can change drastically. Due to this challenge, prior rampdown studies using existing simulators \cite{teplukhina2017simulation, van2023scenarioA, van2023scenarioB, koechl2020evaluation, vincenzi2017eu, asp2022jintrac} typically make assumptions on important effects like the confinement regime transition which is subject to significant uncertainty. These simulation limitations motivated recent experiments at DIII-D designing rampdown trajectories with black box Bayesian Optimization on three control variables, which achieved improvements in the plasma current at time of termination \cite{mehta2024automated}. This experiment showed relatively small adjustments can make an out-sized impact; however, reported pulses, also known as shots, were all at low performance, and a predictive model-based approach is desired for upcoming tokamaks. These limitations motivate the development of models that efficiently learn difficult to simulate dynamics from experimental data, and which are massively parallelizable across uncertainties to enable robust model-based design of trajectories.

To address these challenges, this work takes a data-driven approach, leveraging recent advances made by the Scientific Machine Learning (SciML) community \cite{rackauckas2020universal, kidger2022neural, chen2018neural} and new machine learning frameworks, namely JAX \cite{jax2018github}, which enable the training of dynamics models that combine physics-based equations with  data-driven models. A data-driven approach is not without precedent; aircraft flight control and simulation primarily utilize data-driven models of aerodynamics derived from flight test data in lieu of computational fluid dynamics (CFD) \cite{chyczewski2022summary, allerton2009principles}, often with classical linear state-space models (SSMs) \cite{morelli2016aircraft}. While prior works on learning plasma dynamics using unstructured neural networks required large datasets, often spanning thousands of shots \cite{abbate2021data, char2024full, kit2024learning}, we gain sample efficiency by embedding physics structure into a Neural State Space Model (NSSM) \cite{suykens1995nonlinear, rivals1996black}. This model was trained to generate sufficiently accurate predictions using a modest amount of data, with 311 rampdowns at low performance and only five shots with incomplete rampdowns in the relevant high performance regime, with $\beta_N > 2$ and near the density limit. The model is capable of simulating $\approx 10^4$ rampdown trajectories per second on a single A100 GPU, enabling the usage of the NSSM in a reinforcement learning (RL) training environment. 

The RL environment is massively parallelized to design trajectories with robustness to uncertainties, including the initial conditions of the plasma and its time varying dynamics. We leverage the capabilities of RL for offline design of robust trajectories, which is more readily applicable to the safety-critical settings of upcoming tokamaks than RL for real-time control \cite{degrave2022magnetic, seo2024avoiding, dubbioso2023deep, de2022rl}. A similar approach has previously been demonstrated at KSTAR for designing feed-forward trajectories that reach target states \cite{seo2021feedforward,seo2022development}. After a small number of initial trial shots, the plasma reliably terminated at low plasma current and stored energy for five consecutive high performance shots, with statistically significant improvements relative to baseline, although we encourage caution in interpreting the statistics of the result due to the small sample size. As a test of the viability of this approach for performing small extrapolations in an incremental high performance campaign, which upcoming tokamaks like SPARC and ITER will undergo, we design a rampdown trajectory and perform a predict-first experiment by increasing the plasma current by 20\%, from 140kA to 170kA, for a high $\beta_N$ plasma near the density limit, a scenario for which zero shots of rampdown data exists for TCV. In this extrapolation test, we a priori predict the dynamics of key quantities to within sufficient accuracy to successfully terminate the plasma on both attempts.

The paper is organized as follows. We begin with an overview of the experiment and report the achieved rampdown improvements, as measured by the key figures of merits of plasma current $I_p$ and stored energy $W_{tot}$ at time of plasma termination. Then, an overview of the NSSM is provided along with medium scale validation metrics demonstrating its predictive power. This is followed by an analysis of two shots in the experiment demonstrating the importance of accounting for control errors in trajectory design for preventing a class of disruptions known as vertical displacement events (VDEs). Then, an analysis of 140kA shots in the experiment shows how incremental re-training between run days resulted in rampdowns that are both faster and less disruptive. Results from the predict-first extrapolation test are reported, demonstrating the ability of the NSSM to make small extrapolations. Finally, we discuss future work and implications for upcoming tokamaks like SPARC and ITER.

\section*{Results}
\subsection*{Experiment Overview}
The reported experiment was conducted as a part of the 2024 TCV integrated control, high performance experimental campaign. Flattop plasmas operated at a high performance of $\beta_N > 2.0$ and near the density limit with a highly elongated diverted geometry with $\kappa\approx 1.6$ and $q_{95}\approx 4$. Initial shots in the experiment operated at $I_p=140$ kA, henceforth known as the baseline high performance (HP) scenario, with a final extrapolation test at $I_p=170$ kA. Successful rampdowns from these scenarios require careful management of multiple plasma instability limits that can be exacerbated by details of the plasma trajectories. At present, a comprehensive understanding of disruptive limits remains an open problem, motivating many works on machine-learning based prediction of disruptions \cite{montes2019machine,strait2019progress,zhu2020hybrid,kates2019predicting,vega2022disruption}. However, prior rampdown studies \cite{teplukhina2017simulation,van2023scenarioB,mehta2024automated}, a survey of disruption causes at TCV \cite{labit2024progress}, and the present physics-based understanding of this high density scenario motivated constraints on four plasma parameters. Namely, we impose constraints on the Greenwald fraction, $f_{GW}$, the vertical instability growth rate $\gamma_{vgr}$, the plasma poloidal beta $\beta_p$, and the edge safety factor $q_{95}$. Managing the density limit, which is correlated with the Greenwald fraction, in this scenario is a particular challenge for fast terminations, as the relatively long particle confinement time-scale is a major constraint on the speed of the rampdown. The considered constraints are further discussed in the context of the reward function in the Methods.

\begin{figure}[!htbp]
    \centering
    \includegraphics[width=\linewidth]{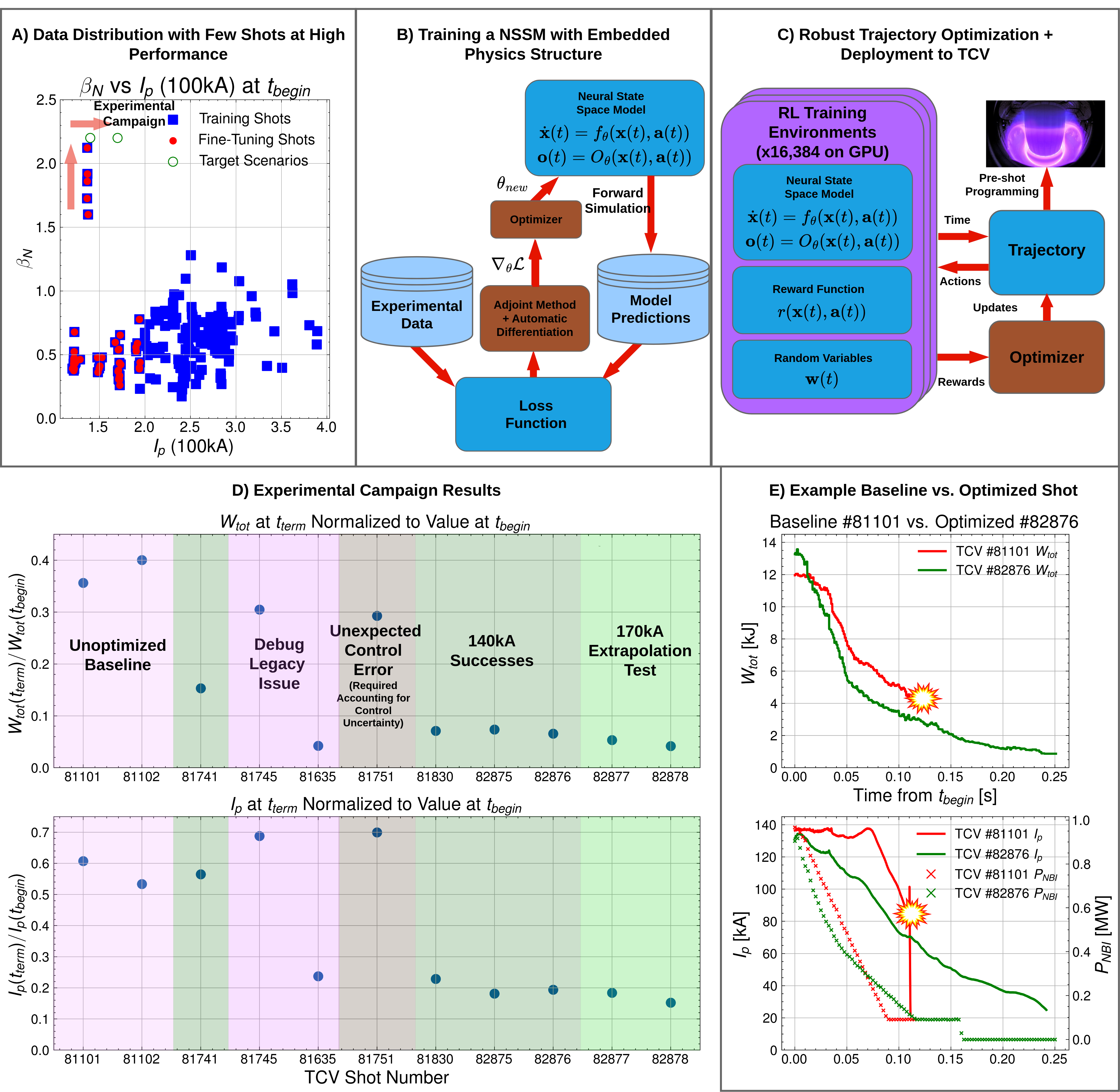}
    \caption{\textbf{Overview of the methodology and key statistical results}. \textbf{a}) The training data distribution, with a modest dataset size at low performance and very few shots in the relevant high performance regime. The target scenarios for this work at 140kA and 170kA high normalized performance are shown. \textbf{b}) Depiction of the dynamics model training method, which involves comparing results from forward simulation of a NSSM against experimental data to compute the gradient of loss with respect to model parameters. \textbf{c}) Depiction of the trajectory optimization process. In addition to the trained dynamics model, the reinforcement learning (RL) training environment is defined by a reward function specifying the desired goal and a set of random variables that training environments are parallelized against, to find a trajectory that has robustness to uncertainties and off-normal events. \textbf{d}) Scatter plot of plasma current, $I_p$, and stored energy, $W_{tot}$, at time of plasma termination. Bottom-right table shows p-values from the Mann-Whitney U test comparing performance of experimental shots, with and without debug shots included, relative to the control set of all shots in the database with $\beta_N$ > 1.5.}
    \label{fig:overview}
\end{figure}

To address the problem, a NSSM dynamics model was trained on a modest dataset of past rampdowns, which contains only 5 incomplete rampdowns in the relevant high performance parameter space, as shown in Figure \ref{fig:overview}A-B. This NSSM is then used in a reinforcement learning (RL) environment to optimize a reward function, designed to minimize time to a goal plasma current of $40$kA and stored energy of $0.5$kJ without disrupting, as shown in Figure \ref{fig:overview}C. The action space was chosen to be plasma current, $I_p$, shaping parameters $\kappa$ and $a_{minor}$ and neutral beam injection (NBI) power $P_{NBI}$. User-specified constraints were set on the Greenwald fraction $f_{GW}$, safety factor $q_{95}$, vertical instability growth rate $\gamma_{vgr}$ as calculated with the method in \cite{marchioni2024vertical}, and poloidal beta $\beta_p$. The optimized action trajectories were then manually programmed into the TCV plasma control system (PCS). The details of the reward function, chosen limits, and PCS programming process are further discussed in the Methods.

Statistical significance of control results in fusion is typically difficult to establish due to the scarcity of experimental time and relevant data-points. This experiment also faces this challenge, given the rampdown experiment involved only nine shots, two of which were dedicated to debugging a legacy software issue, with only five rampdowns in the database near the relevant high performance regime with $\beta_N > 2$. We use these five shots as our control set and define two test sets: one with the debugging shots and one without. As shown in Figure \ref{fig:overview}D, the Mann-Whitney U test \cite{mann1947test} shows a statistically significant improvement in $W_{tot}$ , ($p < 0.05$), at time of plasma termination of the experimental rampdowns for both definitions of the test set. Improvements in $I_p$ are only statistically significant when we do not include the debugging shots. While the results of this statistical test are encouraging, we urge caution in its interpretation given the small sample sizes involved, and the fact that tokamaks are highly drifting distributions in practice, with uncontrolled variables such as wall conditioning making a meaningful impact on experimental outcomes.

Every shot involved in the experiment is shown in Figure 
\ref{fig:shot-breakdown}, showing improvements on $I_p$ and $W_{tot}$ at time of plasma termination over the course of the experimental runs. The un-optimized baseline rampdown trajectory for this scenario disrupted at relatively high current and stored energy in \#81101 and \#81102 at $I_p\approx 80$kA and $W_{tot}\approx 4$kJ. The experiment proceeded iteratively, with re-training of the NSSM on new data and trajectories done after shots \#81635, \#81745, \#81751, and \#81830. A preliminary optimized trajectory was deployed in TCV \#81635, which reached the goal $I_p$ and $W_{tot}$ before disrupting, but post shot analysis showed poor radial control and tracking of the target shape, which was determined to be due to a legacy software issue detailed in Figure S1 in the Supplementary Information. Shots \#81741 and \#81745 were spent resolving this issue, with it properly resolved in \#81751, as shown in Figure S1. \#81751 still disrupted due to a VDE, which was found to be due to a large sensitivity of $\gamma_{vgr}$ to small control errors in the inner gap. After \#81751, an uncertainty distribution on gap errors was added to the RL training environment to gain robustness to this uncertainty, with subsequent shots experiencing similar control errors but without similar increases in $\gamma_{vgr}$, demonstrating the importance of designing trajectories with robustness to real-world uncertainties, as further discussed in Section \ref{subsec:prevent-vde}.

For the final run-day, trajectories were re-optimized, and predictions of the plasma dynamics were generated a priori for both two reprisals of the baseline high performance scenario, but also for the extrapolation test. All four shots for both scenarios terminated successfully below the goal $I_p$, with the baseline scenarios realizing both faster and non-disruptive trajectories relative to baseline and successful a priori predictions of plasma dynamics for both scenarios.

\subsection*{Medium-Scale Validation of NSSM Predictions}
The NSSM was developed and trained to predict the time-dependent dynamics of the set of observations in response to the set of actions listed in Table \ref{tab:obs_action}. The primary goal of the model is to predict the dynamics of key quantities relevant to completing the control task of a fast disruption-free plasma rampdown in response to actuation of controllable variables, to allow the trajectory optimization algorithm to decide on actions that avoid user-specified limits on key quantities correlated with disruptions.

\begin{figure}[!htbp]
    \centering
    \includegraphics[width=0.75\textwidth]{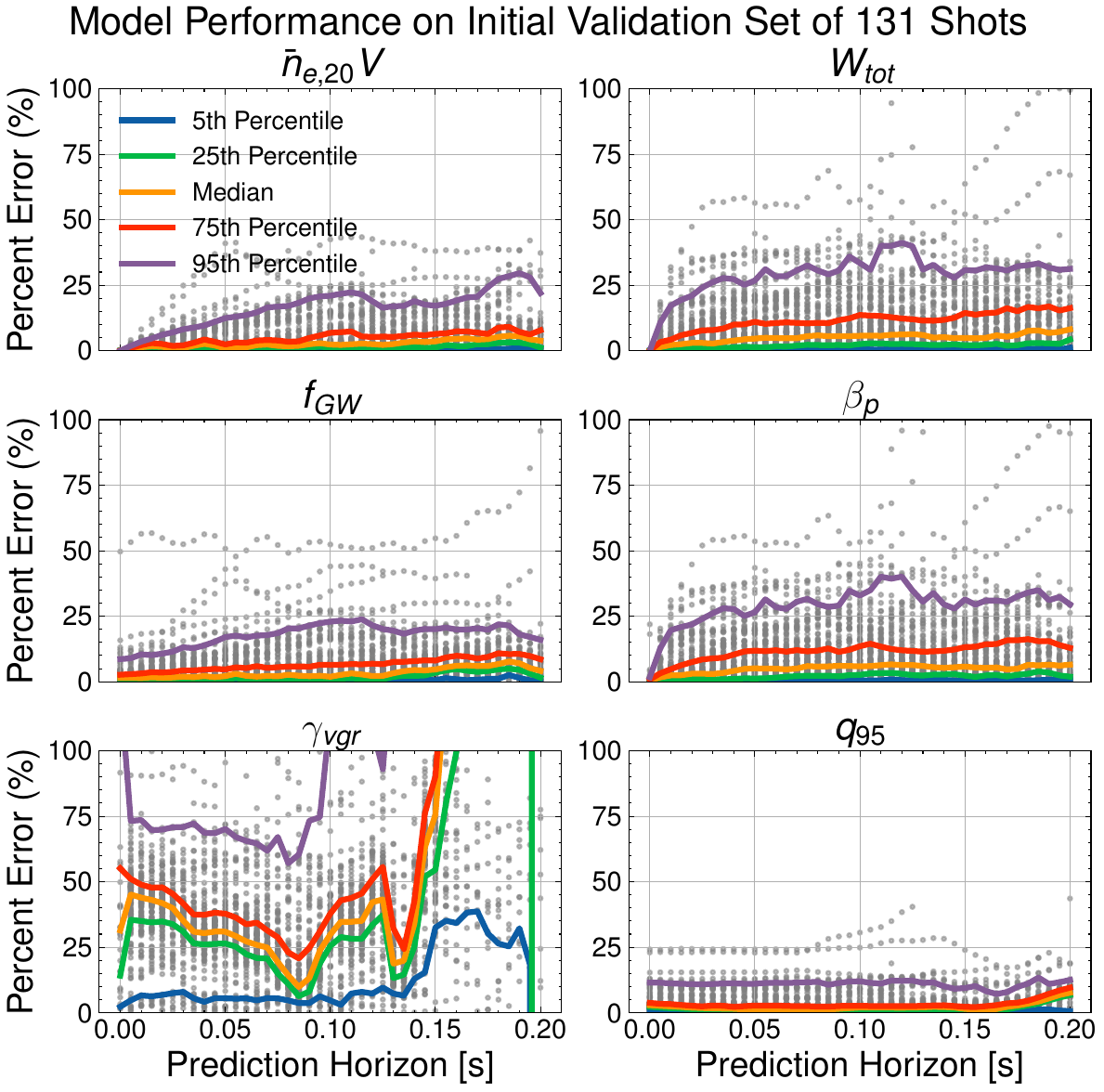}
    \caption{\textbf{Model validation metrics}. Model prediction accuracy as a function of prediction horizon during rampdowns on the validation set of 131 shots. Both individual data points and percentiles are shown. Shown quantities are the line-averaged electron density times volume, $\bar{n}_{e,20}V$, the total stored energy, $W_{tot}$, the Greenwald Fraction, $f_{GW}$, the poloidal beta, $\beta_p$, the vertical instability growth rate, $\gamma_{vgr}$, and the safety factor at the 95th flux surface, $q_{95}$. Figure S4 in the Supplementary Information shows similar model performance on the smaller scale fine tuning dataset.}
    \label{fig:model_val}
\end{figure}

The model underwent two training phases: an initial training phase on a wider dataset with 311 shots in the training dataset and 131 shots in the validation dataset. To improve the predictive power of the model for the relevant scenario, we fine-tune just the confinement scaling of the model by training only on 44 shots with $I_p \leq 200$kA, with all other model weights frozen. Due to the relatively small size of the fine-tuning dataset, we did not separate out a validation or testing dataset for this fine-tuning phase. As shown in Figure \ref{fig:model_val}, the trained model is able to predict the time-dependent dynamics of key 0D kinetic and disruptive quantities to within tens of percent for full rampdowns in the validation dataset, even in the 95th percentile of error. The percent errors for $\gamma_{vgr}$ can be relatively large, but, as shown in Figure S3 in the Supplementary Information, this is largely attributable to the small value of $\gamma_{vgr}$ of limited plasmas as the absolute error is relatively low.

\begin{figure}[!htbp]
    \centering
    \includegraphics[width=\linewidth]{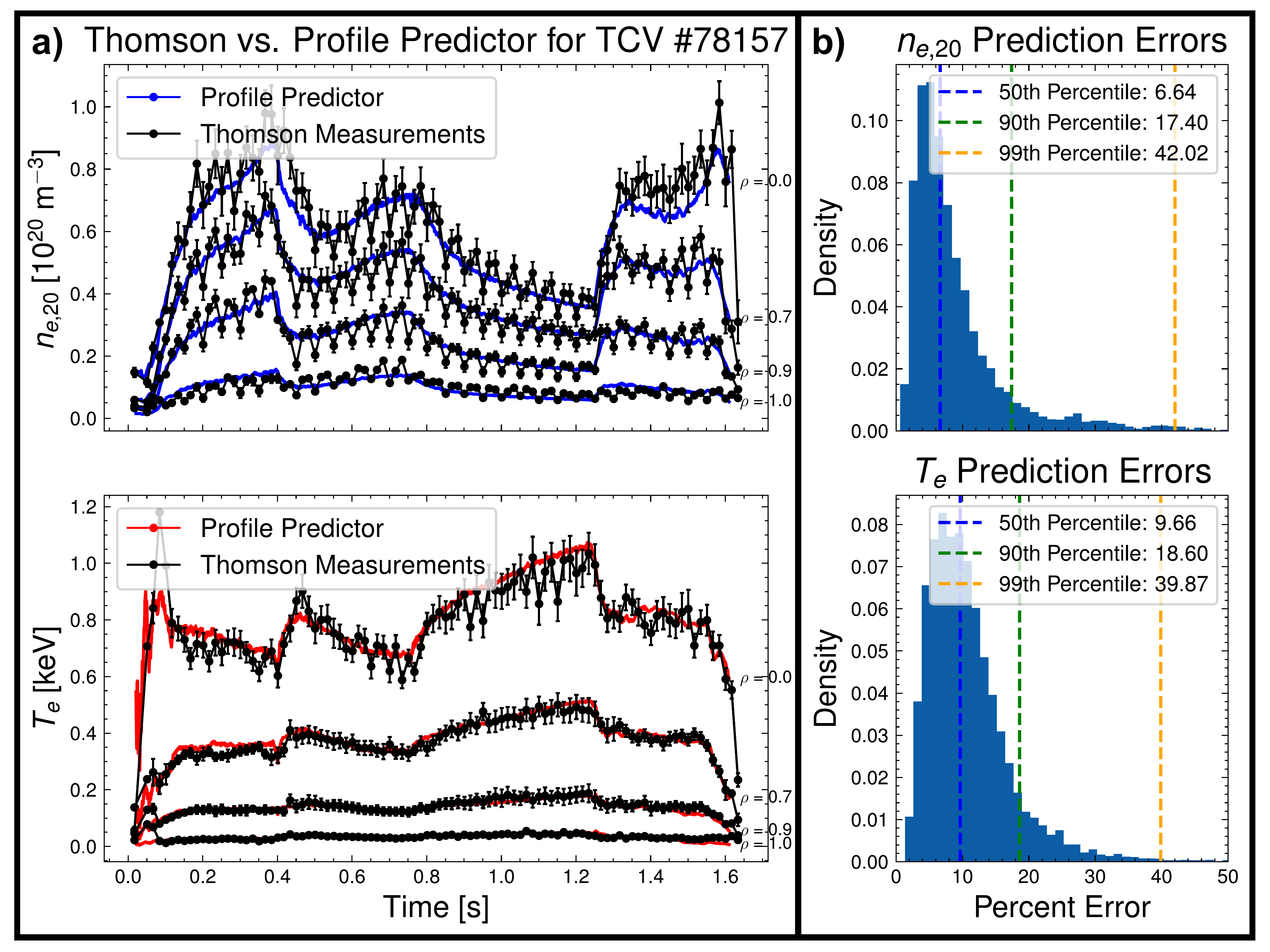}
    \caption{\textbf{Demonstration of profile predictor performance}. \textbf{a)} An example of predictions made by the profile predictor on a validation dataset shot against Thomson measurements, showing both the ability of the model to up-sample Thomson measurements and provide a smoothing effect. Error bars represent two standard deviations. \textbf{b}) The distribution of prediction percent errors on the validation set for electron density, $n_{e,20}$, and electron temperature, $T_e$. The percent error is defined as the integrated error between the prediction and measurement normalized to the average value of the profile. For the $T_e$ profile this is: $100\frac{\int_0^1 |T_{e, Thom}(\rho) - T_{e, pred}(\rho)| d\rho}{\int_0^1 T_{e, Thom}(\rho) d\rho}$. Note that this metric is a pessimistic performance metric as it also captures error due to random measurement noise.}
    \label{fig:prof-predictor-results}
\end{figure}

The NSSM was initially developed with a neural network predictor for the kinetic profiles on the full $\rho$ grid, and initial training runs found that the profile predictor can accurately predict kinetic profiles given the set of 0D scalars specified in Table \ref{tab:obs_action}. Figure \ref{fig:prof-predictor-results} provides an example comparison of predictions of the $T_e$ and $n_e$ profiles against Thomson measurements for a full shot in the validation dataset, showing accurate prediction across all phases of the shot. This result corroborates previous findings at NSTX-U that neural networks can accurately predict kinetic profiles given a set of similar 0D scalars \cite{boyer2021prediction}. Given this result suggests most of the relevant profile information is implicitly captured by 0D scalars, the profile predictor was disabled prior to running experiments to accelerate training, hence reported predictions of kinetic profiles are not predict-first. This result also suggests a structured data-driven approach to modeling tokamak transport merits further research, in parallel with several ongoing principles-based efforts \cite{citrin2024torax,muraca2023reduced,meneghini2024fuse}. Another noteworthy feature of this profile predictor that should be explored in future work is its ability to function as a Thomson up-sampler, as the input variables are all sampled at a higher time resolution, 1ms, than the TCV Thomson Scattering system, which takes measurements every 17ms.

\subsection*{Preventing VDEs by Designing for Robustness to Control Error}
\label{subsec:prevent-vde}
\begin{figure}[!htbp]
    \centering
    \includegraphics[width=0.8\linewidth]{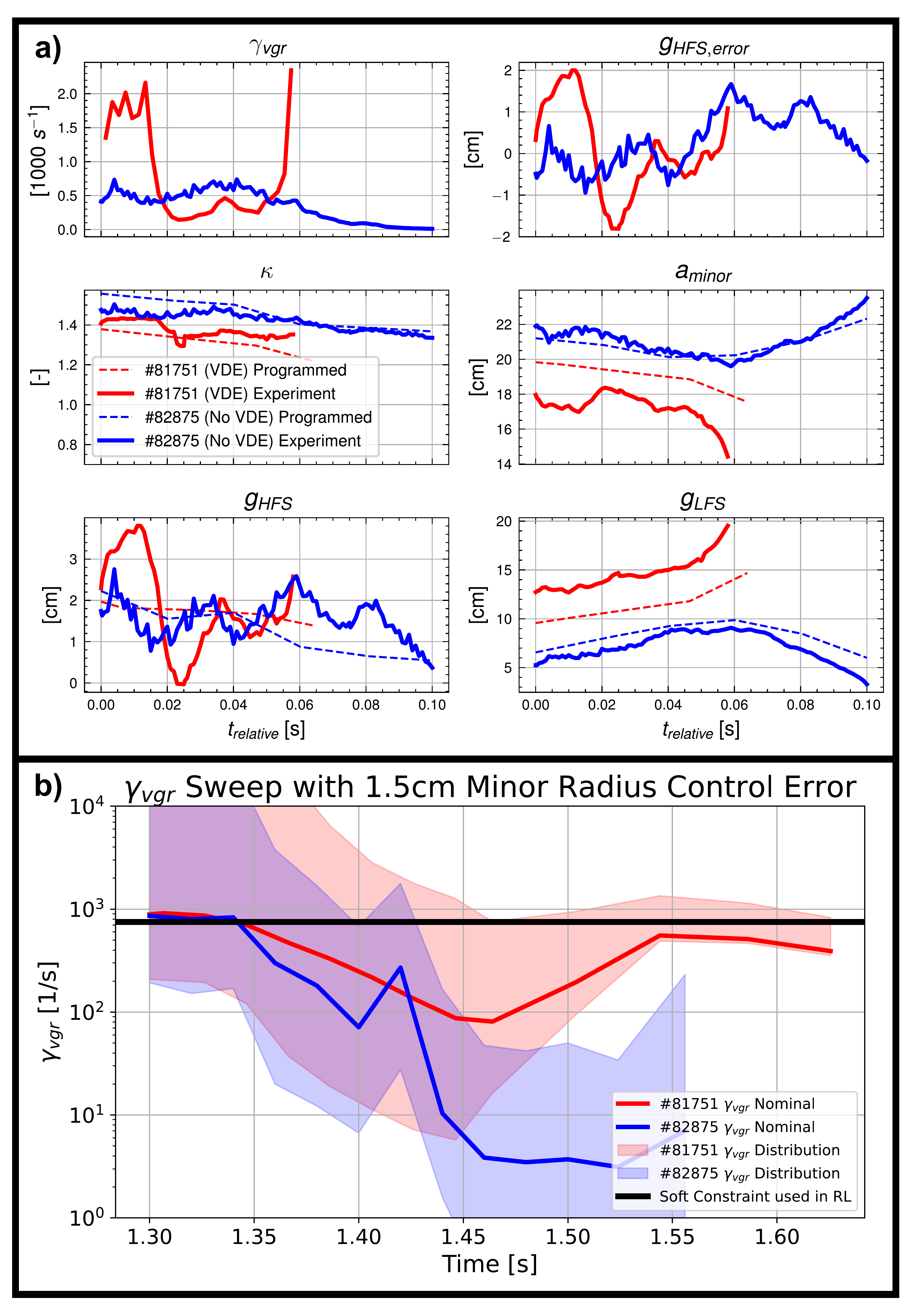}
    \caption{\textbf{Demonstration of increased robustness to control uncertainty}. \textbf{a}) shows a comparison of a shot designed prior to adding control uncertainty, \#81751, and after, \#82875. We see that the vertical instability growth rate, $\gamma_{vgr}$, is highly sensitive to control error in the high field side gap, $g_{HFS,error}$, in \#81751, but similar control errors experienced in \#82875 result in negligible changes in stability. Time shown, $t_{relative}$, is the time relative to 1.4 seconds for \#81751 and 1.3 seconds for \#82875 to align the moment the two shots experience similar control errors. Additional quantities shown are the plasma elongation $\kappa$, minor radius $a_{minor}$, high-field side gap, $g_{HFS}$, and low-field side gap, $g_{LFS}$. \textbf{b}) shows $\gamma_{vgr}$ distributions under 1.5cm of minor radius control error for \#81751 and \#82875. Control error was simulated by introducing variations in the control points provided to the free-boundary equilibrium solver, FBT \cite{hofmann1988fbt,merle2024full}, used for shot preparation. The resulting equilibria were then input into the $\gamma_{vgr}$ computation method used in this work \cite{marchioni2024vertical}.}
    \label{fig:robustness_example}
\end{figure}

The experiment also clearly highlighted the importance of accounting for control errors when optimizing rampdown trajectories. The rampdowns for the initial shots of the experiment were designed without accounting for the impact of uncertainty in shape errors on the vertical growth rate $\gamma_{vgr}$. This uncertainty had a highly sensitive effect in TCV \#81751, which ended in a VDE. Even though the $\gamma_{vgr}$ at zero control error was tolerable, a small increase in the deviation of the high-field-side (HFS) gap, $g_{HFS}$, from the planned value caused an order-of-magnitude increase in $\gamma_{vgr}$, as shown in Figure \ref{fig:robustness_example}.

After \#81751, an uncertainty distribution on the gap errors was added to the RL training environments to encourage the optimization of trajectories that succeed despite this control error. The positive impact of optimizing for robustness to this uncertainty was realized with TCV \#82875, which experienced similar control errors at similar HFS gap values, but without the large increase in $\gamma_{vgr}$. This increased robustness is likely due to a change in the minor radius trajectory, which decreased the low-field-side (LFS) gap, thus increasing the stabilizing effect of the LFS wall whenever the plasma experiences an unexpected outward shift. Prior work at TCV has shown the importance of managing these gaps for vertical stability \cite{marchioni2024vertical}. The importance of accounting for this uncertainty is further highlighted by the fact that \#82875 is more stable in practice than \#81751, despite a higher elongation, the quantity most typically associated with vertical instability. In fact, we can see that \#81751, with its lower elongation, does have a lower $\gamma_{vgr}$ than \#82875 when gap error is near zero, but it is also much more sensitive to control errors. 

The fact that the trajectory in \#82875 is more robust to control error than in \#81751 is corroborated by an analysis using the physics-based model for  $\gamma_{vgr}$ used in this work \cite{marchioni2024vertical}. Minor radius variations were introduced to the RL designed equilibrium trajectories for the two shots, yielding distributions of $\gamma_{vgr}$ trajectories. Figure \ref{fig:robustness_example} shows the conclusion that the trajectories in \#82875 almost uniformly have lower $\gamma_{vgr}$ than \#81751, and stay largely within the soft constraint specified in the RL training process, with the exception of the initial phase of the rampdown process as the flattop equilibrium has a large $\gamma_{vgr}$.

This result demonstrates that the optimal trajectory for minor radius can differ, with significant consequence, once real-world errors and uncertainties are accounted for. Given that existing studies on rampdown design and optimization for ITER \cite{casper2013development} and DEMO \cite{van2023scenarioB}  find solutions involving large reductions in minor radius, these experimental results motivate the further advancement of techniques that enable trajectory design with robustness to uncertainty.

\subsection*{Improving Terminations by Incremental Re-training}
\begin{figure}[!htbp]
    \centering
    \includegraphics[width=0.9\linewidth]{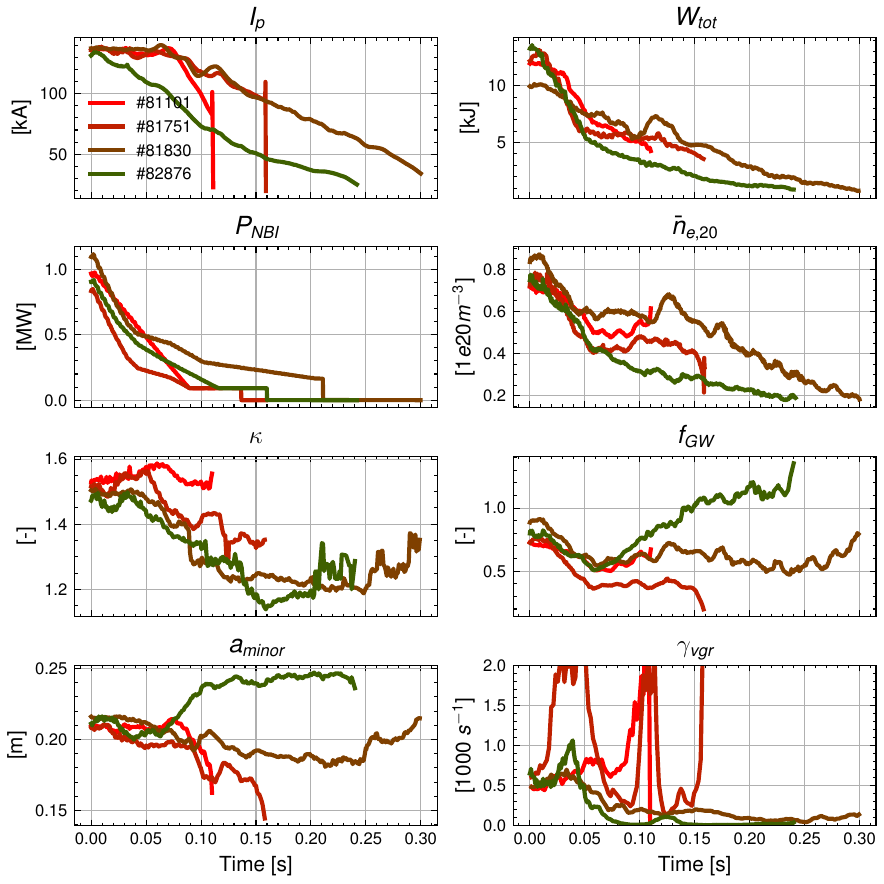}
    \caption{\textbf{Incremental improvements for the baseline high performance scenario}. Experimental traces of key actuators are shown (left), including plasma current, $I_p$, neutral beam power inject, $P_{NBI}$, elongation, $\kappa$, and minor radius, $a_{minor}$. Additional relevant quantities (right) include the stored energy, $W_{tot}$, line-averaged electron density, $\bar{n}_{e,20}$, Greenwald fraction, $f_{GW}$, and vertical instability growth rate, $\gamma_{vgr}$. Time is set relative to the beginning of the termination phase. The plasma current, $I_p$, and stored energy, $W_{tot}$, trajectories show improvement in rampdown speed and disruptivity as the experiment progressed.}
    \label{fig:improve-140}
\end{figure}

Both the NSSM and trajectories were incrementally re-trained on newly generated data from experimental run-days, which resulted in more robust and faster rampdowns for the baseline high performance scenario,  as shown in Figure \ref{fig:improve-140}. The speed of the model enabled re-training of both the model and trajectories in fewer than ten hours total on a single A100 GPU. The un-optimized solution in \#81101 involved a NBI power rampdown while keeping constant plasma current, to allow for a decrease in density to avoid the Greenwald limit, a solution which the RL approach initially decided on as well, as shown with \#81751, with an even more conservative current ramp and introducing a reduction in $\kappa$. As discussed in \ref{subsec:prevent-vde}, this shot resulted in a VDE, and, with the introduction of an uncertainty distribution on $g_{HFS}$, the solution in \#81830 resulted in less of a reduction in the minor radius $a_{minor}$, which helped eliminate the $\gamma_{vgr}$ spikes. Subsequent dynamics model training and trajectory optimization resulted in a solution in \#82876 which allows for an immediate reduction in $I_p$ without running into a density limit, highlighting the ability for the workflow to assist in gradually making improvements. All optimized trajectories involved a fast initial drop in $P_{NBI}$, followed by a slower ramp phase, although the rates and critical points for the transition differed from shot to shot.

\subsection*{Predict-First Results for the Extrapolation Test}\label{subsec:predict-first-170}
Learned dynamics models need not extrapolate far out of distribution to assist with control and trajectory design for net energy tokamaks, as their operations will involve incrementally moving towards higher performance. Thus, they simply need to be able to make reasonable predictions under small extrapolations, and rapidly learn from experiment with as few shots of data as possible. To test the viability of the developed approach in such a setting, we used the learned dynamics model to design trajectories for the extrapolation test scenario, for which zero shots of rampdown data exists in our training dataset for TCV, and generated a priori predictions of the distribution of plasma dynamics during rampdown. 

As shown in Figure \ref{fig:170_predict_first}, experimental results from \#82878 largely fell within this distribution, with accurate predictions of the stored energy and density dynamics. Arguably the largest sources of error came from unreliable control of the plasma shape, contributing to errors in quantities like the rotational transform $\iota_{95}\equiv \frac{1}{q_{95}}$, and also leading to an earlier than expected HL back-transition. \#82878 also started near the density limit, a challenging situation which motivated the introduction of a delay to the $I_p$ ramp in the baseline scenario, but the RL algorithm was able to determine a trajectory to immediately decrease $I_p$, which is desirable, while keeping $f_{GW}$ roughly constant.  \#82877 fell further out of distribution due to a loss of NBI power, and the presence of a neo-classical tearing mode (NTM) at the beginning of rampdown that did not exist in TCV \#82878, as shown by Figure S5 in the Supplementary Information. Fortunately, these un-modeled off-normal events (ONEs) did not take the plasma far enough out of distribution to cause a disruption. As discussed earlier, the profile predictor was removed to help accelerate trajectory optimization, but
post shot evaluation of the profile predictor on the 0D scalars generated by the training environment, shown in Figure \ref{fig:170_predict_first}, shows reasonable agreement against experimental Thomson measurements.

The results from this experiment demonstrate the ability for the learned dynamics model to make small extrapolations to sufficient accuracy to enable the design of robust disruption-free trajectories via RL, and even the prediction misses in TCV \#82877 further emphasize the importance of further advancing the developed methodology to design with robustness to as many ONEs as possible. 

\section*{Discussion}
Our results demonstrate that the developed approach to learning plasma dynamics can predict the highly transient rampdown phase with a modest dataset and even make small extrapolations to higher performance regimes. The relative sample efficiency of the approach, only requiring five shots in the relevant high performance regime, indicates this may be a viable approach for upcoming tokamaks like SPARC and ITER, which will initially operate at low performance before incrementally increasing performance. Developing robust terminations during such incremental campaigns is crucial, as highlighted by the 2020 JET high performance campaign where a 15\% increase in plasma current, from 3MA and 3.5MA, raised disruptivity from $\approx 20\%$ to $\approx 50\%$ \cite{garzotti2023development}. Prediction metrics on the validation dataset, as shown in Figure \ref{fig:model_val}, shows this approach yields accurate predictions for the majority of ramp-downs, but the 5\% worst cases can involve large prediction errors, meriting further investigation.

The developed RL approach for designing robust trajectories yielded promising improvements in the plasma current and stored energy at time of termination, with incremental re-training improving the ramp speed. This result represents one of the first successful demonstrations of trajectory design with robustness to real-world uncertainties for tokamaks, which has historically been infeasible due to the computational cost of simulation. A degree of statistical significance is shown, but the sample size is still relatively small; a larger scale study would more thoroughly determine the efficacy of the approach. Although a large set of uncertainties was accounted for, detailed further in Table S1 in the Supplementary Information, experimental results involved additional uncertainties, such as the NBI failure in \#82877, that still need to be addressed to further improve the robustness of trajectories. Robustness to hardware failure is of particular interest for future work as an exhaustive survey of disruption causes at JET has revealed hardware failure as a significant contributing factor to disruptions \cite{de2011survey}. It is also noteworthy that the RL designed action trajectories tended to be relatively simple, suggesting that the key important ingredient is the fast and parallelized simulation model, as a human expert may be able to find similar trajectories if given access to the simulation model.

To improve the relevance of the developed approach to devices like SPARC and ITER, future work should model additional physics like impurity accumulation and neo-classical tearing mode dynamics, both of which are difficult to simulate, partially stochastic, and have been found to be significant contributing factors to disruptions at JET \cite{de2011survey}.  Accounting for such effects that can drastically change the plasma dynamics may motivate the employment of real-time adjustments to the rampdown trajectory, or the deployment of a library of trajectories as was done in previous simulation studies \cite{wang2024active}. Applying the developed approach to learning JET rampdown dynamics would also further inform the application of this approach to SPARC and ITER.

The developed approach also holds promise for full-shot simulation, which ongoing work is investigating \cite{wang2024plasma}. The ability for a neural network to predict kinetic profiles using 0D scalars, demonstrated both in this work and in prior work \cite{boyer2021prediction}, suggests a data-driven approach may be sufficient for certain control tasks without principles-based transport simulation, which can be extremely computationally expensive and require strong assumptions on edge temperature and density. The ability to deploy accurate, fast, and massively parallel simulators of tokamak plasmas would likely unlock new capabilities for tokamak trajectory and control design, allowing for more reliable access to higher performance plasmas, and ameliorating the risk posed by plasma disruptions to future tokamaks.

\section*{Methods}
\subsection*{The Neural State-Space Model}
\begin{figure}[!htbp]
    \centering
    \includegraphics[width=0.95\linewidth]{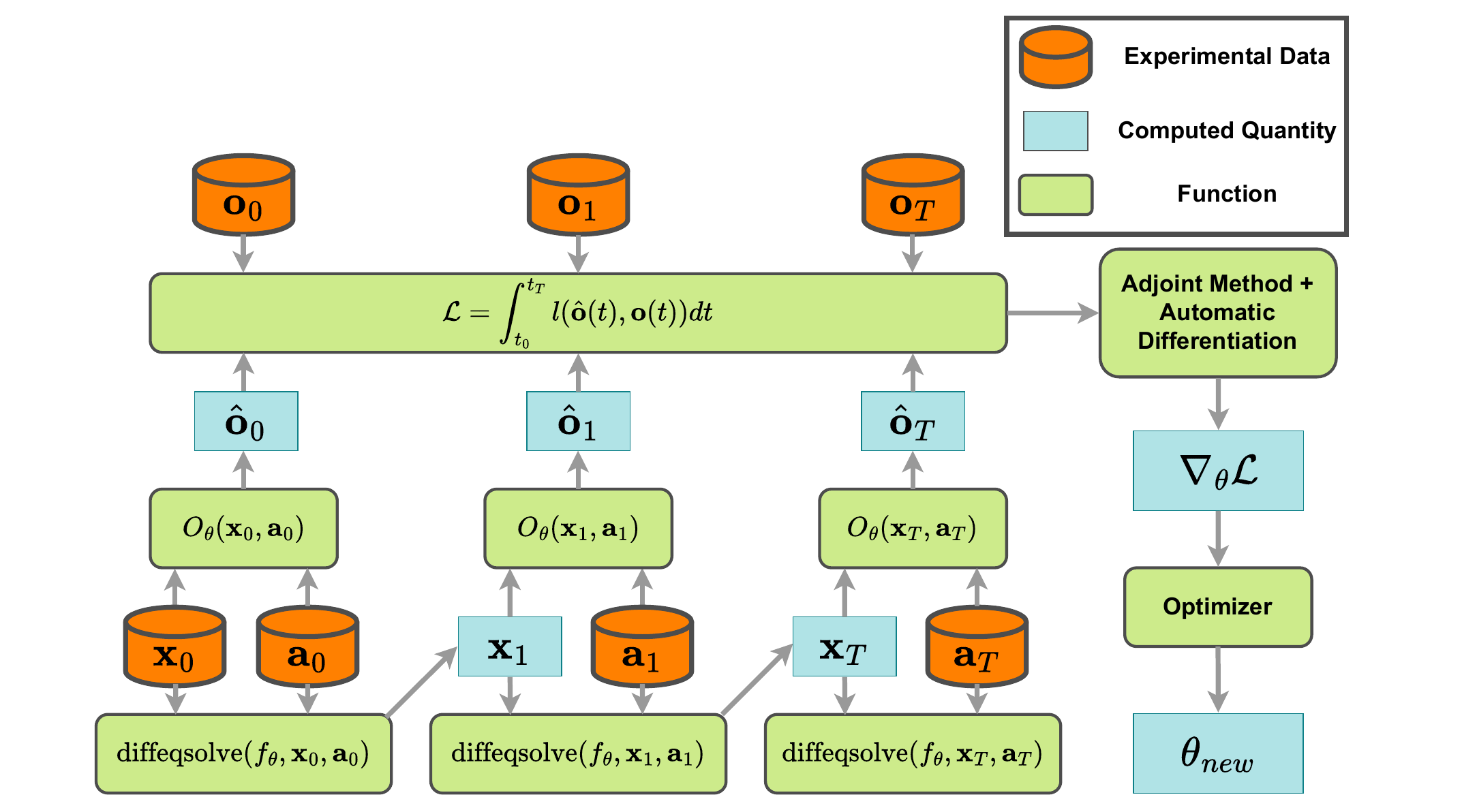}
    \caption{\textbf{Depiction of the Neural State-Space Model (NSSM) training methodology}. The NSSM, defined by the dynamics function $f_\theta$ and observation function $O_\theta$ with parameters $\theta$, is simulated forward in time, given an initial state $\mathbf{x}_0$ and an action trajectory $\mathbf{a}_{0:T}$, to generate a sequence of simulated observations, $\hat{\mathbf{o}}_{0:T}$. The simulated observations are compared with experimental observations via the loss functional $\mathcal{L}$, which is defined as the time-integrated value of an instantaneous loss function $l$. Adjoint methods in diffrax \cite{kidger2022neural} and JAX automatic differentiation then yields the gradient of model parameters with respect to loss, $\nabla_\theta\mathcal{L}$, which allows the optimizer to update the parameters $\theta$.}
    \label{fig:train-nssm}
\end{figure}

Learning dynamical systems from data has been a core discipline within  control design for decades, including aircraft flight control \cite{chyczewski2022summary} and simulation \cite{allerton2009principles}, and has historically been known as system identification\cite{morelli2016aircraft,aastrom1971system}. However, due to computational limitations of the time, classical approaches have typically been restricted to linear models, often in the form of linear state-space models (SSMs):
\begin{subequations}
\begin{align}
        \dot{\mathbf{x}} &= A\mathbf{x} + B\mathbf{a}\\
    \mathbf{o} &= C\mathbf{x} + D\mathbf{a}
\end{align}
\end{subequations}
Where $A$, $B$, $C$, and $D$ are the matrices to be learned from datasets of observables, $\mathbf{o}$, actions, $\mathbf{a}$, and, possibly, states, $\mathbf{x}$. We note that the controls literature typically uses the notation $\mathbf{y}$ in lieu of $\mathbf{o}$ and $\mathbf{u}$ in lieu of $\mathbf{a}$, reflecting a difference in notation between the controls and RL communities, but here we use RL notation for consistency. In the modern deep-learning learning era, this idea of learning dynamical systems from data was re-discovered from a different perspective, with the advent of the neural differential equation (NDE) \cite{chen2018neural}:
\begin{align}
    \dot{\mathbf{x}} = NN_\theta(\mathbf{x})
\end{align}
where it was discovered that, given datasets of $\mathbf{x}$, a neural network, $NN_\theta$, can be used as a system of differential equations that is integrated forward in time with a differential equations solver, and then adjoint back-propagation methods can be used in conjunction with automatic differentiation to determine the gradient of loss with respect to the network parameters $\theta$ \cite{kidger2022neural,chen2018neural,rackauckas2020universal}. The introduction of flexible machine learning frameworks has enabled the development of the field of Scientific Machine Learning (SciML) based around the core idea of extending NDEs to include physics, and other domain-specific, structure \cite{chen2018neural,rackauckas2020universal}. One extension that completes the circle with the classical linear SSM is the neural state-space model (NSSM), which re-introduces the concepts of actions and observations:
\begin{subequations}
    \begin{align}
        \dot{\mathbf{x}}(t) &= f_\theta(\mathbf{x}, \mathbf{a})\\
        \mathbf{o}(t) &= O_\theta(\mathbf{x}, \mathbf{a})
    \end{align}
\end{subequations}
Thanks to the power of new highly flexible machine learning frameworks such as JAX and the Julia SciML ecosystem, $f_\theta$ and $O_\theta$ can be programmed to include arbitrary combinations of neural networks, physics formulas, and even classical data-driven models such as power laws, a capability which we exploit in this work. The training process of a NSSM is shown in Figure \ref{fig:train-nssm}, and involves the simulation of the NSSM forward in time using an initial state $\mathbf{x}_0$ and a time series of actions $\mathbf{a}_{0:T}$ from an experimental database. The error of the simulation results against the experimental data is computed, and adjoint methods and automatic differentiation are used to determine the gradient to reduce the loss. In this work, the differential equation solver package diffrax \cite{kidger2022neural} is used, which includes the integration of multiple adjoint methods with the JAX automatic differentiation system, which allows backpropagation through all differential equation solvers in the package.
\subsubsection*{The Dynamics Function $f_\theta(\mathbf{x}, \mathbf{a})$}\label{subsec:nssm}
We begin by defining the following confinement laws:
\begin{subequations}
\begin{align}
    \tau_{n, pred}(\mathbf{x}, \mathbf{a}) &= c_{n}I_p^{c_{I,n}}\bar{n}_{e,20}^{c_{n,n}}P_{input}^{c_{P,n}}\kappa^{c_{\kappa,n}}\epsilon^{c_{\epsilon,n}}|\dot{I}_p|^{c_{\dot{I}_p,n}}NN_{conf,0}(\mathbf{x}, \mathbf{a})(c_{n,h}\bar{n}_{e,20}^{c_{n,n,h}}P_{input}^{c_{P,n,h}})^{\text{hmode}(\mathbf{x}, \mathbf{a})}\\
    \tau_{E, pred}(\mathbf{x}, \mathbf{a}) &= \underbrace{c_{E}I_p^{c_{I,E}}\bar{n}_{e,20}^{c_{n,E}}P_{input}^{c_{P,E}}\kappa^{c_{\kappa,E}}\epsilon^{c_{\epsilon,E}}|\dot{I}_p|^{c_{\dot{I}_p,E}}}_{\text{L-mode}}\underbrace{NN_{conf,1}(\mathbf{x}, \mathbf{a})}_{\text{NN correction}}\underbrace{(c_{E,h}\bar{n}_{e,20}^{c_{n,E,h}}P_{input}^{c_{P,E,h}})^{\text{hmode}(\mathbf{x}, \mathbf{a})}}_{\text{H-mode correction}}\\
    \text{hmode}(\mathbf{x}, \mathbf{a}) &= \text{tanhHeaviside}(P_{input} - c_{h}\bar{n}_{e,20}^{c_{h,n}}a_{minor}^{c_{h,a}})\\
        \text{tanhClip}(x)&\equiv \tanh\left(\frac{2k}{\text{max} - \text{min}} (x - \text{center})\right) \frac{\text{max} - \text{min}}{2} + \text{center}\\
    \text{tanhHeaviside}(x) &\equiv \frac{1}{2}(\tanh(kx + 1))
\end{align}
\end{subequations}
where the parameters to be learned include all coefficients $c_*$ and neural network parameters. The laws are structured to multiply a portion corresponding to L-mode, a neural network correction factor, and an H-mode correction factor. The L-mode term reflects standard confinement scalings, but with the introduction of a $\dot{I}_p$, which was found to help better capture short-term effects of ramping plasma current. The neural network output includes a tanhClip final activation that constrains it's output to the range $[0.75, 1.25]$, thus controlling the maximum adjustment the network is allowed to make. The $\text{hmode}$ function includes a tanhHeaviside function which provides a smooth transition between one to zero once $P_{input}$ falls below the learned back-transition threshold, which is structured to reflect the Martin scaling \cite{martin2008power}. Note that the use of the $\text{hmode}$ function output as a power de-activates the H-mode correction term once $\text{hmode}$ transitions from one to zero. While, in principle, the neural network should be able to learn the effects of H-mode implicitly, we found that adding an explicit H-mode correction factor helped improve model predictions in our low-data regime. The $k$ factor controls the smoothness of both the tanhClip and tanhHeaviside functions.

These confinement laws are integrated as a part of the following 0D energy and particle balance equations, which is a model that blends simple physics principles, power laws, and neural networks:
\begin{subequations}
    \begin{align}
    \frac{dW_{tot}}{dt} &= -\underbrace{\frac{W_{tot}}{\tau_{E, pred}}}_{\text{Transport}} + \underbrace{I_p^2NN_{ohm, rad, 0}(\mathbf{x}, \mathbf{a})}_{\text{Ohmic Heating}} - \underbrace{\bar{n}_{e, 20}VNN_{ohm, rad, 1}(\mathbf{x}, \mathbf{a})}_{\text{Radiated Power}} + \underbrace{P_{NBI} + P_{ECRH}}_{\text{Aux. Heating}}\\
    \frac{d(\bar{n}_{e,20}V)}{dt} &= -\underbrace{\frac{\bar{n}_{e, 20}}{\tau_{n, pred}}}_{\text{Transport}} + \underbrace{c_{NBI}P_{NBI}}_{\text{NBI Fueling}} + \underbrace{c_{gas,0}\sigma(c_{gas,1}V_{gas} + c_{gas,2})}_{\text{Gas Valve Fueling}} + \underbrace{NN_{wall}(\mathbf{x}, \mathbf{a})\exp^{-c_{wall}g_{HFS}}}_{\text{Wall Effects}}
\end{align}
\end{subequations}
where $\sigma$ is the sigmoid function, $NN_{ohm, rad}$ is a network that predicts two quantities; the first is multiplied by $I_p^2$ to serve as an Ohmic heating term, and the second is multiplied by density and volume to serve as the radiated power term. $NN_{wall}$ is included to account for possible wall fueling effects when in a limited configuration, and is multiplied by an exponential in the HFS gap to de-activate it when diverted. Additional simple constants are included to account for fueling from both NBI and gas puffing. We note that, in both cases, the included terms do not capture important state dependencies and time delays, but they proved sufficient for this use case. The dynamics of density times volume are predicted; in cases where density itself is used (e.g. to compute the Greenwald Fraction), the following volume approximation is used to recover density:
\begin{align}
    V\approx 2\pi R^2\epsilon^2\kappa \left(\pi - \left(\pi - \frac{8}{3}\right)\epsilon\right)
\end{align}
Since time derivatives of quantities, $\dot{I}_p, \dot{\kappa}, \dot{a}_{minor}, \dot{\delta}$ are used as actions, their integrated values are also added as state variables with trivial dynamics.

\subsubsection*{The Observation Function $O_\theta(\mathbf{x}, \mathbf{a})$}
The observation function consists of several components: a NN predictor for $\gamma_{vgr}$, a profile predictor, and simple physics formulae to compute derived quantities:
\begin{subnumcases}{O_\theta(\mathbf{x}, \mathbf{a})}
        \beta_p = \frac{8}{3}\frac{W_{tot}}{\mu_0R_0I_p^2}& \\
    f_{GW} = \frac{\bar{n}_{e, 20}\pi a_{minor}^2}{I_{p,MA}}&\\
    q_{95} = \frac{4.1 a_{minor}^2 B_0}{R_0  I_{p, MA}} 
 \left(1 + 1.2  (\kappa - 1) + 0.56 (\kappa - 1)^2\right) 
 \left(1 + 0.09  \delta + 0.16  \delta^2\right) 
 \frac{1 + 0.45  \delta  \epsilon}{1 - 0.74  \epsilon}&\\
    \gamma_{vgr} = NN_{vgr}(\mathbf{x}, \mathbf{a})&\\
    T_e(\rho), N_e(\rho) = NN_{prof}(\mathbf{x}, \mathbf{a})
\end{subnumcases}
where $\beta_p$ is computed in accordance to the LIUQE definition \cite{moret2015tokamak}, $f_{GW}$ is the usual Greenwald Fraction \cite{greenwald1988new}, $q_{95}$ is the approximation given in \cite{sauter2016geometric} with the squareness factor set to 1, $NN_{vgr}$ is a multi-layer perceptron (MLP with GELU activation and a scaled sigmoid output, and $NN_{prof}$ is a neural-operator based profile predictor, discussed further in the next subsection.

\subsubsection*{Neural Operator Based Profile Predictor}
 Prior work at NSTX-U trained a neural network to successfully predict kinetic profile shapes using their averages plus zero dimensional control parameters such as plasma current, shaping, and auxiliary heating. Building upon this prior work, we show that, on TCV data, kinetic profiles can be predicted to reasonable accuracy with a neural network using the stored energy $W_{tot}$, line-averaged electron density $\bar{n}_{e, 20}$, and control parameters. The key implication is that accurate prediction of the time-dependent dynamics of just two scalars, $W_{tot}$ and $\bar{n}_{e, 20}$, implies reasonable prediction of the dynamics of kinetic profiles.

We leverage methods developed by the neural operator \cite{kovachki2023neural, li2020neural} literature, which has found success for solving machine learning problems in scientific domains involving PDEs. Letting $f_{in}$ denote an input function and $f_{out}$ denote an output function, a neural operator $\mathcal{F}$ parameterized by $\theta$ maps an input function to an output function:
\begin{align}
    f_{out} = \mathcal{F}_\theta(f_{in})
\end{align}
In practice, the functions involved are approximated using a set of basis functions, thus the practical implementation results in a neural network operating on basis function coefficients. In this work, we make use of cubic B-spline basis functions to represent the kinetic profiles:
\begin{align}
    T_{e}(\rho) = \sum_{i=1}^{n_{basis}}c_{T,i}B_{i, 3}(\rho) \quad n_{e, 20}(\rho) = \sum_{i=1}^{n_{basis}}c_{n, i}B_{i, 3}(\rho)
\end{align}
And we predict these profiles using a set of 0D scalars, where every scalar is a control parameter except stored energy $W_{tot}$ and $\bar{n}_{e,20}$. The full set of input and output parameters is specified in Table \ref{tab:obs_action}. During training, the $\rho$ grid corresponding to the dataset is chosen to evaluate the basis functions, but arbitrary alternative grids can be used during inference time.

\subsubsection*{Training Methods}
Training of the NSSM involved two stages. First, $NN_{ohm,rad}$, $NN_{vgr}$, and $NN_{prof}$ are trained independently of the rest of the model on time-independent samples to predict their respective quantities. These ``pretrained'' models are then integrated into the NSSM, where they are further trained jointly with the rest of the model through the time-dependent process specified in Figure \ref{fig:train-nssm}. The AdamW optimizer \cite{loshchilov2017decoupled} with an exponential decay learning rate schedule is used for every training run . All NNs used in the dynamics function $f$ and $NN_{vgr}$ are simple multi-layer perceptrons with GELU \cite{hendrycks2016gaussian} activations on the hidden state and tanhClip functions as final activations to constrain their outputs to reasonable ranges. The profile predictor is further detailed earlier in the methods. Hyperparameters for the optimizer and model sizes are optimized via Bayesian Optimization using the method implemented in the Weights and Biases platform \cite{wandb}, which was used in this work for experiment tracking. The final set of hyperparameters are detailed in Table S2 in the Supplementary Information.

\subsection*{Training Data Distribution}
The dataset used for training models in this work consists of the 442 most recent shots with rampdowns that are at least partially complete and sufficient diagnostic availability, gathered with DEFUSE (Disruption and Event analysis framework for FUSion Experiments) \cite{pau2023modern}. The initial training phase involved training on 311 shots of data, with the rest of the dataset used for validation. After the initial training phase, the model is further trained on a fine-tuning dataset of 44 shots, during this phase all of the model weights except those in the $\tau_E$ and $\tau_N$ hybrid confinement laws described in \ref{subsec:nssm} are frozen. As shown in Figure S8 in the Supplementary Information, the dataset consists of only five shots of data anywhere near the relevant high-performance region.

\subsection*{Reward Function}
The reward function is designed to balance the priority of achieving a low plasma current and energy against the risk of disrupting the plasma, and is given by:
\begin{equation} \label{eq:reward_fn}
    r(\mathbf{x}(t), \mathbf{a}(t)) = \underbrace{-c_{time}}_{\text{Penalty for time}} - \underbrace{c_{W}W_{tot}(t) - c_{I_p}I_{p}(t)}_{\text{Penalty for current and energy}} - \underbrace{\sum_{i=1}^{n_{soft}} c_{soft}s_i(\mathbf{x}(t))}_{\text{Soft chance-constraints}} - \underbrace{\sum_{i=1}^{n_{hard}} c_{hard}h_i(\mathbf{x}(t))}_{\text{Hard chance-constraints}}
\end{equation}
The reward function is active for every time step before hitting the goal state or maximum allowed training episode time. The goal state is chosen to be a stored energy of 500J and a plasma current of 40kA, as, for the 170kA extrapolation test scenario, 40kA approximately corresponds to the relative plasma current for an ITER 15MA benign termination, which is defined as 3MA \cite{de2017multi}. A constant penalty term is active for every time step before achieving the goal to encourage time minimization. In addition, penalty terms that scale with plasma current and energy are included to further prioritize moving towards a safer state. To avoid disruptive limits such as high Greenwald fraction during the rampdown, penalty terms are added for states that violate user-specified constraints on key quantities correlated with disruptions. 

One challenge with specifying constraint limits is the difference in severity of violating different constraints, and the, at times, weak correlations between physical quantities and disruptions. To address this issue, we partition constraints into ``soft'' constraints, which incur a small penalty to discourage, but not forbid, the algorithm from finding solutions that violates these limits, and ``hard'' constraints which incur a large penalty to strictly enforce constraint violation. We note that while methods in the constrained optimization literature often mathematically express constraints separately from the objective function being optimized, most practical implementations of constrained optimization algorithms enforce constraints by rewriting constraints as penalty terms in the objective function \cite{nesterov2018lectures,bertsekas2014constrained}, an approach we also adopt. Stochastic optimization across a distribution of outcomes introduces a challenge: trying to avoid limits for every scenario will likely result in excessively slow and conservative solutions \cite{bertsimas2004price}, which itself poses it's own risk. To address this challenge, we utilize chance-constraints, a technique often utilized in the autonomous driving literature \cite{qin2023review, wang2020non}, and only activate the constraint if violation probability exceeds a certain threshold. In this set of experiments, this threshold is chosen as 5\%. Reward function parameters used for the final four shots are shown in Table S2 in the Supplementary Information.

\subsection*{Uncertainty Model}\label{subsec:uncertainty_model}
In experimental reality, the time evolution of plasma dynamics is highly nonlinear and subject to considerable amounts of uncertainty, as evidenced by the two same-scenario shots shown in Figure \ref{fig:170_predict_first} which begin at drastically different initial conditions. To design trajectories that have robustness to large variability and off-normal events, we defined an uncertainty model for the RL training environments, and sampled from this uncertainty model for each training environment used during training. The uncertainty model includes random variables for both the initial state of the plasma during rampdown and disturbances/model uncertainties that affect the time-varying plasma dynamics. To account for the fact that accidental H-L back-transition implies the initial state of the plasma may start in either H or L-mode, the initial state distribution is modeled as a bi-modal mixture model, with a 50\% chance of any given RL training environment starting in either H or L-mode. In some cases, uncertainty distributions could easily be quantified from past experimental data (such as tracking error in the plasma current), or from model prediction accuracy (such as $\gamma_{vgr}$), but in other cases the distribution was chosen in ad-hoc fashion, upon identifying additional sources of uncertainty in experiment. Table S1 in the Supplementary Information summarizes the random variables, parameterized distributions, and quantification methods used in this work. As discussed in Section \ref{subsec:predict-first-170}, this uncertainty model proved to be non-exhaustive in experiment. In addition, the uncertainty model employed does not account for time-varying fluctuations in uncertain variables; future work should employ time-varying stochastic processes. Both of these limitations further highlight the need to advance experimental uncertainty quantification and robust control in the context of fusion plasma control.

\subsection*{RL Methods}
Standard RL problems involve optimizing a policy $\pi$ to map observations to actions:
\begin{align}
    \mathbf{a} = \pi(\mathbf{o})
\end{align}
from this perspective, trajectory optimization can be viewed as policy optimization where the only observable is time:
\begin{align}
    \mathbf{a} = \pi(t)
\end{align}
Given that time is the only observable, but there exists different physical conditions in the parallel training environments which are unobservable to the policy, the reward maximization process yields a trajectory that is designed to succeed across the different conditions specified in Subsection \ref{subsec:uncertainty_model}. After an initial trial with Proximal Policy Optimization (PPO) \cite{schulman2017proximal}, we found OpenAI-ES, an evolutionary strategy (ES) designed for policy optimization, to work better in practice \cite{salimans2017evolution}. This is possibly explained by the theoretical analysis given in the paper introducing OpenAI-ES, which suggests that RL problems with long time horizons and actions that have long-lasting effects may be better solved with ES approaches than the dominant paradigm of policy gradient methods \cite{salimans2017evolution}. The policy $\pi$ was parameterized by a multi-layer perceptron with two hidden layers of width 64 and used relu activations with a hyperbolic tangent final activation to constrain the action space. Hyperparameter sweep of the architecture was not employed and would be worthwhile to investigate for future works.
\subsection*{Deployment to TCV}
Shape trajectories determined via RL were mapped to last-closed-flux-surface control points via re-scaling of the flat-top shape for the diverted phase, and using an analytic formula in the TCV MGAMS \cite{hofmann1994creation} algorithm for the limited phase. Feed-forward coil currents and voltages to achieve the desired plasma current and shaping trajectories were then determined with the free-boundary equilibrium code FBT and shot preparation algorithm MGAMS \cite{hofmann1994creation, hofmann1988fbt}, and the $P_{NBI}$ trajectory was programmed into the TCV supervisory control system SAMONE \cite{galperti2024overview,vu2021integrated}.

\section*{Data Availability}
Data to generate figures found in this paper are available in the source code repository at https://doi.org/10.5281/zenodo.16621120. The complete dataset used for training and validation can be obtained by contacting Alessandro Pau or Allen M. Wang.
\section*{Code Availability}
Source code is provided at https://doi.org/10.5281/zenodo.16621120.

\bibliography{references}

\begin{thebibliography}{10}
\urlstyle{rm}
\expandafter\ifx\csname url\endcsname\relax
  \def\url#1{\texttt{#1}}\fi
\expandafter\ifx\csname urlprefix\endcsname\relax\def\urlprefix{URL }\fi
\expandafter\ifx\csname doiprefix\endcsname\relax\def\doiprefix{DOI: }\fi
\providecommand{\bibinfo}[2]{#2}
\providecommand{\eprint}[2][]{\url{#2}}

\bibitem{duval2024experimental}
\bibinfo{author}{Duval, B.} \emph{et~al.}
\newblock \bibinfo{journal}{\bibinfo{title}{Experimental research on the tcv tokamak}}.
\newblock {\emph{\JournalTitle{Nuclear Fusion}}} \textbf{\bibinfo{volume}{64}}, \bibinfo{pages}{112023} (\bibinfo{year}{2024}).

\bibitem{creely2020overview}
\bibinfo{author}{Creely, A.} \emph{et~al.}
\newblock \bibinfo{journal}{\bibinfo{title}{Overview of the sparc tokamak}}.
\newblock {\emph{\JournalTitle{Journal of Plasma Physics}}} \textbf{\bibinfo{volume}{86}}, \bibinfo{pages}{865860502} (\bibinfo{year}{2020}).

\bibitem{shimada2007overview}
\bibinfo{author}{Shimada, M.} \emph{et~al.}
\newblock \bibinfo{journal}{\bibinfo{title}{Overview and summary}}.
\newblock {\emph{\JournalTitle{Nuclear Fusion}}} \textbf{\bibinfo{volume}{47}}, \bibinfo{pages}{S1} (\bibinfo{year}{2007}).

\bibitem{maris2023impact}
\bibinfo{author}{Maris, A.~D.}, \bibinfo{author}{Wang, A.}, \bibinfo{author}{Rea, C.}, \bibinfo{author}{Granetz, R.} \& \bibinfo{author}{Marmar, E.}
\newblock \bibinfo{journal}{\bibinfo{title}{The impact of disruptions on the economics of a tokamak power plant}}.
\newblock {\emph{\JournalTitle{Fusion Science and Technology}}} \bibinfo{pages}{1--17} (\bibinfo{year}{2023}).

\bibitem{sorbom2015arc}
\bibinfo{author}{Sorbom, B.} \emph{et~al.}
\newblock \bibinfo{journal}{\bibinfo{title}{Arc: A compact, high-field, fusion nuclear science facility and demonstration power plant with demountable magnets}}.
\newblock {\emph{\JournalTitle{Fusion Engineering and Design}}} \textbf{\bibinfo{volume}{100}}, \bibinfo{pages}{378--405} (\bibinfo{year}{2015}).

\bibitem{federici2014overview}
\bibinfo{author}{Federici, G.} \emph{et~al.}
\newblock \bibinfo{journal}{\bibinfo{title}{Overview of eu demo design and r\&d activities}}.
\newblock {\emph{\JournalTitle{Fusion Engineering and Design}}} \textbf{\bibinfo{volume}{89}}, \bibinfo{pages}{882--889} (\bibinfo{year}{2014}).

\bibitem{van2023scenarioA}
\bibinfo{author}{Van~Mulders, S.} \emph{et~al.}
\newblock \bibinfo{journal}{\bibinfo{title}{Scenario optimization for the tokamak ramp-down phase in raptor. part a: Analysis and model validation on asdex upgrade}}.
\newblock {\emph{\JournalTitle{Plasma Physics and Controlled Fusion}}}  (\bibinfo{year}{2023}).

\bibitem{teplukhina2017simulation}
\bibinfo{author}{Teplukhina, A.} \emph{et~al.}
\newblock \bibinfo{journal}{\bibinfo{title}{Simulation of profile evolution from ramp-up to ramp-down and optimization of tokamak plasma termination with the raptor code}}.
\newblock {\emph{\JournalTitle{Plasma Physics and Controlled Fusion}}} \textbf{\bibinfo{volume}{59}}, \bibinfo{pages}{124004} (\bibinfo{year}{2017}).

\bibitem{mehta2024automated}
\bibinfo{author}{Mehta, V.} \emph{et~al.}
\newblock \bibinfo{journal}{\bibinfo{title}{Automated experimental design of safe rampdowns via probabilistic machine learning}}.
\newblock {\emph{\JournalTitle{Nuclear Fusion}}} \textbf{\bibinfo{volume}{64}}, \bibinfo{pages}{046014} (\bibinfo{year}{2024}).

\bibitem{greenwald1988new}
\bibinfo{author}{Greenwald, M.} \emph{et~al.}
\newblock \bibinfo{journal}{\bibinfo{title}{A new look at density limits in tokamaks}}.
\newblock {\emph{\JournalTitle{Nuclear Fusion}}} \textbf{\bibinfo{volume}{28}}, \bibinfo{pages}{2199} (\bibinfo{year}{1988}).

\bibitem{troyon1984mhd}
\bibinfo{author}{Troyon, F.}, \bibinfo{author}{Gruber, R.}, \bibinfo{author}{Saurenmann, H.}, \bibinfo{author}{Semenzato, S.} \& \bibinfo{author}{Succi, S.}
\newblock \bibinfo{journal}{\bibinfo{title}{Mhd-limits to plasma confinement}}.
\newblock {\emph{\JournalTitle{Plasma Physics and Controlled Fusion}}} \textbf{\bibinfo{volume}{26}}, \bibinfo{pages}{209} (\bibinfo{year}{1984}).

\bibitem{garzotti2023development}
\bibinfo{author}{Garzotti, L.} \emph{et~al.}
\newblock \bibinfo{journal}{\bibinfo{title}{Development of baseline scenario for high fusion performance at jet, submitted to}}.
\newblock {\emph{\JournalTitle{Nuclear Fusion}}}  (\bibinfo{year}{2023}).

\bibitem{van2023scenarioB}
\bibinfo{author}{Van~Mulders, S.} \emph{et~al.}
\newblock \bibinfo{journal}{\bibinfo{title}{Scenario optimization for the tokamak ramp-down phase in raptor. part b: Safe termination of demo plasmas}}.
\newblock {\emph{\JournalTitle{Plasma Physics and Controlled Fusion}}}  (\bibinfo{year}{2023}).

\bibitem{koechl2020evaluation}
\bibinfo{author}{Koechl, F.} \emph{et~al.}
\newblock \bibinfo{journal}{\bibinfo{title}{Evaluation of fuelling requirements for core density and divertor heat load control in non-stationary phases of the iter dt 15 ma baseline scenario}}.
\newblock {\emph{\JournalTitle{Nuclear Fusion}}} \textbf{\bibinfo{volume}{60}}, \bibinfo{pages}{066015} (\bibinfo{year}{2020}).

\bibitem{vincenzi2017eu}
\bibinfo{author}{Vincenzi, P.} \emph{et~al.}
\newblock \bibinfo{journal}{\bibinfo{title}{Eu demo transient phases: main constraints and heating mix studies for ramp-up and ramp-down}}.
\newblock {\emph{\JournalTitle{Fusion Engineering and Design}}} \textbf{\bibinfo{volume}{123}}, \bibinfo{pages}{473--476} (\bibinfo{year}{2017}).

\bibitem{asp2022jintrac}
\bibinfo{author}{Asp, E.~M.} \emph{et~al.}
\newblock \bibinfo{journal}{\bibinfo{title}{Jintrac integrated simulations of iter scenarios including fuelling and divertor power flux control for h, he and dt plasmas}}.
\newblock {\emph{\JournalTitle{Nuclear Fusion}}} \textbf{\bibinfo{volume}{62}}, \bibinfo{pages}{126033} (\bibinfo{year}{2022}).

\bibitem{rackauckas2020universal}
\bibinfo{author}{Rackauckas, C.} \emph{et~al.}
\newblock \bibinfo{journal}{\bibinfo{title}{Universal differential equations for scientific machine learning}}.
\newblock {\emph{\JournalTitle{arXiv preprint arXiv:2001.04385}}}  (\bibinfo{year}{2020}).

\bibitem{kidger2022neural}
\bibinfo{author}{Kidger, P.}
\newblock \bibinfo{journal}{\bibinfo{title}{On neural differential equations}}.
\newblock {\emph{\JournalTitle{arXiv preprint arXiv:2202.02435}}}  (\bibinfo{year}{2022}).

\bibitem{chen2018neural}
\bibinfo{author}{Chen, R.~T.}, \bibinfo{author}{Rubanova, Y.}, \bibinfo{author}{Bettencourt, J.} \& \bibinfo{author}{Duvenaud, D.~K.}
\newblock \bibinfo{journal}{\bibinfo{title}{Neural ordinary differential equations}}.
\newblock {\emph{\JournalTitle{Advances in neural information processing systems}}} \textbf{\bibinfo{volume}{31}} (\bibinfo{year}{2018}).

\bibitem{jax2018github}
\bibinfo{author}{Bradbury, J.} \emph{et~al.}
\newblock \bibinfo{title}{{JAX}: composable transformations of {P}ython+{N}um{P}y programs} (\bibinfo{year}{2018}).

\bibitem{chyczewski2022summary}
\bibinfo{author}{Chyczewski, T.~S.}, \bibinfo{author}{Lofthouse, A.~J.}, \bibinfo{author}{Gea, L.-M.}, \bibinfo{author}{Cartieri, A.} \& \bibinfo{author}{Hiller, B.~R.}
\newblock \bibinfo{title}{Summary of the first aiaa stability and control prediction workshop}.
\newblock In \emph{\bibinfo{booktitle}{AIAA SciTech 2022 Forum}}, \bibinfo{pages}{1680} (\bibinfo{year}{2022}).

\bibitem{allerton2009principles}
\bibinfo{author}{Allerton, D.}
\newblock \emph{\bibinfo{title}{Principles of flight simulation}}, vol.~\bibinfo{volume}{27} (\bibinfo{publisher}{John Wiley \& Sons}, \bibinfo{year}{2009}).

\bibitem{morelli2016aircraft}
\bibinfo{author}{Morelli, E.~A.} \& \bibinfo{author}{Klein, V.}
\newblock \emph{\bibinfo{title}{Aircraft system identification: theory and practice}}, vol.~\bibinfo{volume}{2} (\bibinfo{publisher}{Sunflyte Enterprises Williamsburg, VA}, \bibinfo{year}{2016}).

\bibitem{abbate2021data}
\bibinfo{author}{Abbate, J.}, \bibinfo{author}{Conlin, R.} \& \bibinfo{author}{Kolemen, E.}
\newblock \bibinfo{journal}{\bibinfo{title}{Data-driven profile prediction for diii-d}}.
\newblock {\emph{\JournalTitle{Nuclear Fusion}}} \textbf{\bibinfo{volume}{61}}, \bibinfo{pages}{046027} (\bibinfo{year}{2021}).

\bibitem{char2024full}
\bibinfo{author}{Char, I.}, \bibinfo{author}{Chung, Y.}, \bibinfo{author}{Abbate, J.}, \bibinfo{author}{Kolemen, E.} \& \bibinfo{author}{Schneider, J.}
\newblock \bibinfo{journal}{\bibinfo{title}{Full shot predictions for the diii-d tokamak via deep recurrent networks}}.
\newblock {\emph{\JournalTitle{arXiv preprint arXiv:2404.12416}}}  (\bibinfo{year}{2024}).

\bibitem{kit2024learning}
\bibinfo{author}{Kit, A.} \emph{et~al.}
\newblock \bibinfo{journal}{\bibinfo{title}{On learning latent dynamics of the aug plasma state}}.
\newblock {\emph{\JournalTitle{Physics of Plasmas}}} \textbf{\bibinfo{volume}{31}} (\bibinfo{year}{2024}).

\bibitem{suykens1995nonlinear}
\bibinfo{author}{Suykens, J.~A.}, \bibinfo{author}{De~Moor, B.~L.} \& \bibinfo{author}{Vandewalle, J.}
\newblock \bibinfo{journal}{\bibinfo{title}{Nonlinear system identification using neural state space models, applicable to robust control design}}.
\newblock {\emph{\JournalTitle{International Journal of Control}}} \textbf{\bibinfo{volume}{62}}, \bibinfo{pages}{129--152} (\bibinfo{year}{1995}).

\bibitem{rivals1996black}
\bibinfo{author}{Rivals, I.} \& \bibinfo{author}{Personnaz, L.}
\newblock \bibinfo{title}{Black-box modeling with state-space neural networks}.
\newblock In \emph{\bibinfo{booktitle}{Neural Adaptive Control Technology}}, \bibinfo{pages}{237--264} (\bibinfo{publisher}{World Scientific}, \bibinfo{year}{1996}).

\bibitem{degrave2022magnetic}
\bibinfo{author}{Degrave, J.} \emph{et~al.}
\newblock \bibinfo{journal}{\bibinfo{title}{Magnetic control of tokamak plasmas through deep reinforcement learning}}.
\newblock {\emph{\JournalTitle{Nature}}} \textbf{\bibinfo{volume}{602}}, \bibinfo{pages}{414--419} (\bibinfo{year}{2022}).

\bibitem{seo2024avoiding}
\bibinfo{author}{Seo, J.} \emph{et~al.}
\newblock \bibinfo{journal}{\bibinfo{title}{Avoiding fusion plasma tearing instability with deep reinforcement learning}}.
\newblock {\emph{\JournalTitle{Nature}}} \textbf{\bibinfo{volume}{626}}, \bibinfo{pages}{746--751} (\bibinfo{year}{2024}).

\bibitem{dubbioso2023deep}
\bibinfo{author}{Dubbioso, S.} \emph{et~al.}
\newblock \bibinfo{journal}{\bibinfo{title}{A deep reinforcement learning approach for vertical stabilization of tokamak plasmas}}.
\newblock {\emph{\JournalTitle{Fusion Engineering and Design}}} \textbf{\bibinfo{volume}{194}}, \bibinfo{pages}{113725} (\bibinfo{year}{2023}).

\bibitem{de2022rl}
\bibinfo{author}{De~Tommasi, G.} \emph{et~al.}
\newblock \bibinfo{title}{A rl-based vertical stabilization system for the east tokamak}.
\newblock In \emph{\bibinfo{booktitle}{2022 American Control Conference (ACC)}}, \bibinfo{pages}{5328--5333} (\bibinfo{organization}{IEEE}, \bibinfo{year}{2022}).

\bibitem{seo2021feedforward}
\bibinfo{author}{Seo, J.} \emph{et~al.}
\newblock \bibinfo{journal}{\bibinfo{title}{Feedforward beta control in the kstar tokamak by deep reinforcement learning}}.
\newblock {\emph{\JournalTitle{Nuclear Fusion}}} \textbf{\bibinfo{volume}{61}}, \bibinfo{pages}{106010} (\bibinfo{year}{2021}).

\bibitem{seo2022development}
\bibinfo{author}{Seo, J.} \emph{et~al.}
\newblock \bibinfo{journal}{\bibinfo{title}{Development of an operation trajectory design algorithm for control of multiple 0d parameters using deep reinforcement learning in kstar}}.
\newblock {\emph{\JournalTitle{Nuclear Fusion}}} \textbf{\bibinfo{volume}{62}}, \bibinfo{pages}{086049} (\bibinfo{year}{2022}).

\bibitem{montes2019machine}
\bibinfo{author}{Montes, K.~J.} \emph{et~al.}
\newblock \bibinfo{journal}{\bibinfo{title}{Machine learning for disruption warnings on alcator c-mod, diii-d, and east}}.
\newblock {\emph{\JournalTitle{Nuclear Fusion}}} \textbf{\bibinfo{volume}{59}}, \bibinfo{pages}{096015} (\bibinfo{year}{2019}).

\bibitem{strait2019progress}
\bibinfo{author}{Strait, E.} \emph{et~al.}
\newblock \bibinfo{journal}{\bibinfo{title}{Progress in disruption prevention for iter}}.
\newblock {\emph{\JournalTitle{Nuclear Fusion}}} \textbf{\bibinfo{volume}{59}}, \bibinfo{pages}{112012} (\bibinfo{year}{2019}).

\bibitem{zhu2020hybrid}
\bibinfo{author}{Zhu, J.} \emph{et~al.}
\newblock \bibinfo{journal}{\bibinfo{title}{Hybrid deep-learning architecture for general disruption prediction across multiple tokamaks}}.
\newblock {\emph{\JournalTitle{Nuclear Fusion}}} \textbf{\bibinfo{volume}{61}}, \bibinfo{pages}{026007} (\bibinfo{year}{2020}).

\bibitem{kates2019predicting}
\bibinfo{author}{Kates-Harbeck, J.}, \bibinfo{author}{Svyatkovskiy, A.} \& \bibinfo{author}{Tang, W.}
\newblock \bibinfo{journal}{\bibinfo{title}{Predicting disruptive instabilities in controlled fusion plasmas through deep learning}}.
\newblock {\emph{\JournalTitle{Nature}}} \textbf{\bibinfo{volume}{568}}, \bibinfo{pages}{526--531} (\bibinfo{year}{2019}).

\bibitem{vega2022disruption}
\bibinfo{author}{Vega, J.}, \bibinfo{author}{Murari, A.}, \bibinfo{author}{Dormido-Canto, S.}, \bibinfo{author}{Ratt{\'a}, G.~A.} \& \bibinfo{author}{Gelfusa, M.}
\newblock \bibinfo{journal}{\bibinfo{title}{Disruption prediction with artificial intelligence techniques in tokamak plasmas}}.
\newblock {\emph{\JournalTitle{Nature Physics}}} \textbf{\bibinfo{volume}{18}}, \bibinfo{pages}{741--750} (\bibinfo{year}{2022}).

\bibitem{labit2024progress}
\bibinfo{author}{Labit, B.} \emph{et~al.}
\newblock \bibinfo{journal}{\bibinfo{title}{Progress in the development of the iter baseline scenario in tcv}}.
\newblock {\emph{\JournalTitle{Plasma Physics and Controlled Fusion}}} \textbf{\bibinfo{volume}{66}}, \bibinfo{pages}{025016} (\bibinfo{year}{2024}).

\bibitem{marchioni2024vertical}
\bibinfo{author}{Marchioni, S.}
\newblock \emph{\bibinfo{title}{Vertical Instability Studies in the TCV Tokamak and Development and Application of Multimachine Real-Time Proximity Control Strategies}}.
\newblock Ph.D. thesis, \bibinfo{school}{EPFL} (\bibinfo{year}{2024}).
\newblock \doiprefix\url{10.5075/epfl-thesis-10943}.

\bibitem{mann1947test}
\bibinfo{author}{Mann, H.~B.} \& \bibinfo{author}{Whitney, D.~R.}
\newblock \bibinfo{journal}{\bibinfo{title}{On a test of whether one of two random variables is stochastically larger than the other}}.
\newblock {\emph{\JournalTitle{The annals of mathematical statistics}}} \bibinfo{pages}{50--60} (\bibinfo{year}{1947}).

\bibitem{boyer2021prediction}
\bibinfo{author}{Boyer, M.~D.} \& \bibinfo{author}{Chadwick, J.}
\newblock \bibinfo{journal}{\bibinfo{title}{Prediction of electron density and pressure profile shapes on nstx-u using neural networks}}.
\newblock {\emph{\JournalTitle{Nuclear Fusion}}} \textbf{\bibinfo{volume}{61}}, \bibinfo{pages}{046024} (\bibinfo{year}{2021}).

\bibitem{citrin2024torax}
\bibinfo{author}{Citrin, J.} \emph{et~al.}
\newblock \bibinfo{journal}{\bibinfo{title}{Torax: A fast and differentiable tokamak transport simulator in jax}}.
\newblock {\emph{\JournalTitle{arXiv preprint arXiv:2406.06718}}}  (\bibinfo{year}{2024}).

\bibitem{muraca2023reduced}
\bibinfo{author}{Muraca, M.} \emph{et~al.}
\newblock \bibinfo{journal}{\bibinfo{title}{Reduced transport models for a tokamak flight simulator}}.
\newblock {\emph{\JournalTitle{Plasma Physics and Controlled Fusion}}} \textbf{\bibinfo{volume}{65}}, \bibinfo{pages}{035007} (\bibinfo{year}{2023}).

\bibitem{meneghini2024fuse}
\bibinfo{author}{Meneghini, O.} \emph{et~al.}
\newblock \bibinfo{journal}{\bibinfo{title}{Fuse (fusion synthesis engine): A next generation framework for integrated design of fusion pilot plants}}.
\newblock {\emph{\JournalTitle{arXiv preprint arXiv:2409.05894}}}  (\bibinfo{year}{2024}).

\bibitem{casper2013development}
\bibinfo{author}{Casper, T.} \emph{et~al.}
\newblock \bibinfo{journal}{\bibinfo{title}{Development of the iter baseline inductive scenario}}.
\newblock {\emph{\JournalTitle{Nuclear Fusion}}} \textbf{\bibinfo{volume}{54}}, \bibinfo{pages}{013005} (\bibinfo{year}{2013}).

\bibitem{de2011survey}
\bibinfo{author}{De~Vries, P.} \emph{et~al.}
\newblock \bibinfo{journal}{\bibinfo{title}{Survey of disruption causes at jet}}.
\newblock {\emph{\JournalTitle{Nuclear fusion}}} \textbf{\bibinfo{volume}{51}}, \bibinfo{pages}{053018} (\bibinfo{year}{2011}).

\bibitem{wang2024active}
\bibinfo{author}{Wang, A.~M.} \emph{et~al.}
\newblock \bibinfo{journal}{\bibinfo{title}{Active disruption avoidance and trajectory design for tokamak ramp-downs with neural differential equations and reinforcement learning}}.
\newblock {\emph{\JournalTitle{arXiv preprint arXiv:2402.09387}}}  (\bibinfo{year}{2024}).

\bibitem{wang2024plasma}
\bibinfo{author}{Wang, A.} \emph{et~al.}
\newblock \bibinfo{journal}{\bibinfo{title}{Plasma operational simulation (popsim): A control-oriented simulation toolbox for parallel simulation, system identification, and optimization}}.
\newblock {\emph{\JournalTitle{Bulletin of the American Physical Society}}}  (\bibinfo{year}{2024}).

\bibitem{aastrom1971system}
\bibinfo{author}{{\AA}str{\"o}m, K.~J.} \& \bibinfo{author}{Eykhoff, P.}
\newblock \bibinfo{journal}{\bibinfo{title}{System identification—a survey}}.
\newblock {\emph{\JournalTitle{Automatica}}} \textbf{\bibinfo{volume}{7}}, \bibinfo{pages}{123--162} (\bibinfo{year}{1971}).

\bibitem{martin2008power}
\bibinfo{author}{Martin, Y.}, \bibinfo{author}{Takizuka, T.} \emph{et~al.}
\newblock \bibinfo{title}{Power requirement for accessing the h-mode in iter}.
\newblock In \emph{\bibinfo{booktitle}{Journal of Physics: Conference Series}}, vol. \bibinfo{volume}{123}, \bibinfo{pages}{012033} (\bibinfo{organization}{IOP Publishing}, \bibinfo{year}{2008}).

\bibitem{moret2015tokamak}
\bibinfo{author}{Moret, J.-M.} \emph{et~al.}
\newblock \bibinfo{journal}{\bibinfo{title}{Tokamak equilibrium reconstruction code liuqe and its real time implementation}}.
\newblock {\emph{\JournalTitle{Fusion Engineering and Design}}} \textbf{\bibinfo{volume}{91}}, \bibinfo{pages}{1--15} (\bibinfo{year}{2015}).

\bibitem{sauter2016geometric}
\bibinfo{author}{Sauter, O.}
\newblock \bibinfo{journal}{\bibinfo{title}{Geometric formulas for system codes including the effect of negative triangularity}}.
\newblock {\emph{\JournalTitle{Fusion Engineering and Design}}} \textbf{\bibinfo{volume}{112}}, \bibinfo{pages}{633--645} (\bibinfo{year}{2016}).

\bibitem{kovachki2023neural}
\bibinfo{author}{Kovachki, N.} \emph{et~al.}
\newblock \bibinfo{journal}{\bibinfo{title}{Neural operator: Learning maps between function spaces with applications to pdes}}.
\newblock {\emph{\JournalTitle{Journal of Machine Learning Research}}} \textbf{\bibinfo{volume}{24}}, \bibinfo{pages}{1--97} (\bibinfo{year}{2023}).

\bibitem{li2020neural}
\bibinfo{author}{Li, Z.} \emph{et~al.}
\newblock \bibinfo{journal}{\bibinfo{title}{Neural operator: Graph kernel network for partial differential equations}}.
\newblock {\emph{\JournalTitle{arXiv preprint arXiv:2003.03485}}}  (\bibinfo{year}{2020}).

\bibitem{loshchilov2017decoupled}
\bibinfo{author}{Loshchilov, I.} \& \bibinfo{author}{Hutter, F.}
\newblock \bibinfo{journal}{\bibinfo{title}{Decoupled weight decay regularization}}.
\newblock {\emph{\JournalTitle{arXiv preprint arXiv:1711.05101}}}  (\bibinfo{year}{2017}).

\bibitem{hendrycks2016gaussian}
\bibinfo{author}{Hendrycks, D.} \& \bibinfo{author}{Gimpel, K.}
\newblock \bibinfo{journal}{\bibinfo{title}{Gaussian error linear units (gelus)}}.
\newblock {\emph{\JournalTitle{arXiv preprint arXiv:1606.08415}}}  (\bibinfo{year}{2016}).

\bibitem{wandb}
\bibinfo{author}{Biewald, L.}
\newblock \bibinfo{title}{Experiment tracking with weights and biases} (\bibinfo{year}{2020}).
\newblock \bibinfo{note}{Software available from wandb.com}.

\bibitem{pau2023modern}
\bibinfo{author}{Pau, A.} \emph{et~al.}
\newblock \bibinfo{title}{A modern framework to support disruption studies: the eurofusion disruption database}.
\newblock In \emph{\bibinfo{booktitle}{29th IAEA Int. Conf. on Fusion Energy (London, UK, 2023)}}, \bibinfo{pages}{p--EX} (\bibinfo{organization}{IAEA}, \bibinfo{year}{2023}).

\bibitem{de2017multi}
\bibinfo{author}{de~Vries, P.~C.} \emph{et~al.}
\newblock \bibinfo{journal}{\bibinfo{title}{Multi-machine analysis of termination scenarios with comparison to simulations of controlled shutdown of iter discharges}}.
\newblock {\emph{\JournalTitle{Nuclear Fusion}}} \textbf{\bibinfo{volume}{58}}, \bibinfo{pages}{026019} (\bibinfo{year}{2017}).

\bibitem{nesterov2018lectures}
\bibinfo{author}{Nesterov, Y.} \emph{et~al.}
\newblock \emph{\bibinfo{title}{Lectures on convex optimization}}, vol. \bibinfo{volume}{137} (\bibinfo{publisher}{Springer}, \bibinfo{year}{2018}).

\bibitem{bertsekas2014constrained}
\bibinfo{author}{Bertsekas, D.~P.}
\newblock \emph{\bibinfo{title}{Constrained optimization and Lagrange multiplier methods}} (\bibinfo{publisher}{Academic press}, \bibinfo{year}{2014}).

\bibitem{bertsimas2004price}
\bibinfo{author}{Bertsimas, D.} \& \bibinfo{author}{Sim, M.}
\newblock \bibinfo{journal}{\bibinfo{title}{The price of robustness}}.
\newblock {\emph{\JournalTitle{Operations research}}} \textbf{\bibinfo{volume}{52}}, \bibinfo{pages}{35--53} (\bibinfo{year}{2004}).

\bibitem{qin2023review}
\bibinfo{author}{Qin, H.} \emph{et~al.}
\newblock \bibinfo{journal}{\bibinfo{title}{Review of autonomous path planning algorithms for mobile robots}}.
\newblock {\emph{\JournalTitle{Drones}}} \textbf{\bibinfo{volume}{7}}, \bibinfo{pages}{211} (\bibinfo{year}{2023}).

\bibitem{wang2020non}
\bibinfo{author}{Wang, A.}, \bibinfo{author}{Jasour, A.} \& \bibinfo{author}{Williams, B.~C.}
\newblock \bibinfo{journal}{\bibinfo{title}{Non-gaussian chance-constrained trajectory planning for autonomous vehicles under agent uncertainty}}.
\newblock {\emph{\JournalTitle{IEEE Robotics and Automation Letters}}} \textbf{\bibinfo{volume}{5}}, \bibinfo{pages}{6041--6048} (\bibinfo{year}{2020}).

\bibitem{schulman2017proximal}
\bibinfo{author}{Schulman, J.}, \bibinfo{author}{Wolski, F.}, \bibinfo{author}{Dhariwal, P.}, \bibinfo{author}{Radford, A.} \& \bibinfo{author}{Klimov, O.}
\newblock \bibinfo{journal}{\bibinfo{title}{Proximal policy optimization algorithms}}.
\newblock {\emph{\JournalTitle{arXiv preprint arXiv:1707.06347}}}  (\bibinfo{year}{2017}).

\bibitem{salimans2017evolution}
\bibinfo{author}{Salimans, T.}, \bibinfo{author}{Ho, J.}, \bibinfo{author}{Chen, X.}, \bibinfo{author}{Sidor, S.} \& \bibinfo{author}{Sutskever, I.}
\newblock \bibinfo{journal}{\bibinfo{title}{Evolution strategies as a scalable alternative to reinforcement learning}}.
\newblock {\emph{\JournalTitle{arXiv preprint arXiv:1703.03864}}}  (\bibinfo{year}{2017}).

\bibitem{hofmann1994creation}
\bibinfo{author}{Hofmann, F.} \emph{et~al.}
\newblock \bibinfo{journal}{\bibinfo{title}{Creation and control of variably shaped plasmas in tcv}}.
\newblock {\emph{\JournalTitle{Plasma Physics and Controlled Fusion}}} \textbf{\bibinfo{volume}{36}}, \bibinfo{pages}{B277} (\bibinfo{year}{1994}).

\bibitem{hofmann1988fbt}
\bibinfo{author}{Hofmann, F.}
\newblock \bibinfo{journal}{\bibinfo{title}{Fbt-a free-boundary tokamak equilibrium code for highly elongated and shaped plasmas}}.
\newblock {\emph{\JournalTitle{Computer Physics Communications}}} \textbf{\bibinfo{volume}{48}}, \bibinfo{pages}{207--221} (\bibinfo{year}{1988}).

\bibitem{galperti2024overview}
\bibinfo{author}{Galperti, C.} \emph{et~al.}
\newblock \bibinfo{journal}{\bibinfo{title}{Overview of the tcv digital real-time plasma control system and its applications}}.
\newblock {\emph{\JournalTitle{Fusion Engineering and Design}}} \textbf{\bibinfo{volume}{208}}, \bibinfo{pages}{114640} (\bibinfo{year}{2024}).

\bibitem{vu2021integrated}
\bibinfo{author}{Vu, T.} \emph{et~al.}
\newblock \bibinfo{journal}{\bibinfo{title}{Integrated real-time supervisory management for off-normal-event handling and feedback control of tokamak plasmas}}.
\newblock {\emph{\JournalTitle{IEEE Transactions on Nuclear Science}}} \textbf{\bibinfo{volume}{68}}, \bibinfo{pages}{1855--1861} (\bibinfo{year}{2021}).

\bibitem{giacomin2022first}
\bibinfo{author}{Giacomin, M.} \emph{et~al.}
\newblock \bibinfo{journal}{\bibinfo{title}{First-principles density limit scaling in tokamaks based on edge turbulent transport and implications for iter}}.
\newblock {\emph{\JournalTitle{Physical Review Letters}}} \textbf{\bibinfo{volume}{128}}, \bibinfo{pages}{185003} (\bibinfo{year}{2022}).

\bibitem{maris2024correlation}
\bibinfo{author}{Maris, A.~D.} \emph{et~al.}
\newblock \bibinfo{journal}{\bibinfo{title}{Correlation of the l-mode density limit with edge collisionality}}.
\newblock {\emph{\JournalTitle{Nuclear Fusion}}} \textbf{\bibinfo{volume}{65}}, \bibinfo{pages}{016051} (\bibinfo{year}{2024}).

\bibitem{merle2024full}
\bibinfo{author}{Merle, A.}, \bibinfo{author}{Felici, F.}, \bibinfo{author}{Heiss, C.}, \bibinfo{author}{Van~Parys, G.} \& \bibinfo{author}{Wai, J.}
\newblock \bibinfo{title}{Full discharge coil trajectory optimisation using a quasi-newton method with the fbt code from the meq suite}.
\newblock In \emph{\bibinfo{booktitle}{50th EPS Conference on Controlled Fusion and Plasma Physics}} (\bibinfo{year}{2024}).

\end{thebibliography}

\section*{Acknowledgments}
Allen M. Wang and Cristina Rea were funded in part by Commonwealth Fusion Systems. This work has been carried out within the frame-work of the EUROfusion Consortium, via the Euratom Research and Training Programme (Grant Agreement No 101052200 - EUROfusion) and funded by the Swiss State Secretariat for Education, Research, and Innovation (SERI). Alessandro Pau, Olivier Sauter, Anna Vu, Cristian Galperti, Antoine Merle, Yoeri Poels, Cristina Venturini, Federico Felici, and Stefano Marchioni received support from EUROfusion and SERI. Views and opinions expressed are however those of the author(s) only and do not necessarily reflect those of the European Union, the European Commission, or SERI. Neither the European Union nor the European Commission nor SERI can be held responsible for them. The authors would like to acknowledge the Engaging cluster, managed by the MIT Office of Research Computing and Data, which was used in this work for model training.

\section*{Author Contributions Statement}
Allen M. Wang led the project, developed the dynamics modeling approach and RL problem formulation, and led writing of the paper. Alessandro Pau developed the machine learning dataset, high performance integrated control scenario, and integration of the approach into TCV. Oswin So and Charles Dawson worked with Allen M. Wang to develop the RL training environments and methods. Olivier Sauter enabled TCV integration, and identified and debugged the radial observer issue. Cristina Rea managed the collaboration, advised the project on disruptions and machine learning, and contributed to manuscript drafting and revisions. Mark D. Boyer advised the project from a controls perspective and motivated the approach taken for dynamics modeling. Anna Vu and Cristian Galperti enabled integration with the integrated control system. Chuchu Fan advised Allen M. Wang, Oswin So, and Charles Dawson on controls and RL approaches. Antoine Merle developed and assisted with MHD equilibrium codes. Yoeri Poels and Cristina Venturini helped with the TCV dataset. Federico Felici developed equilibrium and plasma control infrastructure that enabled this work and advised development of the growth rate calculation method. Stefano Marchioni developed the growth rate calculation method used in this work.

\section*{Competing Interests Statement}
Mark Dan Boyer is an employee of Commonwealth Fusion Systems. Allen M. Wang and Cristina Rea are funded in part by Commonwealth Fusion Systems.
\newpage
\section*{Tables}
\begin{table}[!htbp]
\centering
\renewcommand{\arraystretch}{1.2} 
\begin{tabularx}{\textwidth}{@{}lXc@{}} 
\toprule
\textbf{Parameter} & \textbf{Description} & \textbf{Units} \\
\midrule
\multicolumn{3}{c}{\textbf{Actions} $\mathbf{a}_t$} \\
\addlinespace
$dI_p/dt$ & Plasma current ramp-rate & MA/s \\
$P_{NBI}$ & Neutral beam injection (NBI) heating & MW \\
$d\kappa/dt$ & Elongation ramp-rate & 1/s \\
$da_{minor}/dt$ & Minor radius ramp-rate & m/s \\
$d\delta/dt$ & Triangularity ramp-rate (zeroed for trajectory optimization) & 1/s\\
$g_{HFS}$ & Gap between the separatrix and the high-field side wall (constant 0.02m and zeroed after NBI  off) & m \\
$P_{ECRH}$ & Electron cyclotron resonance heating (ECRH) injected power (zeroed for trajectory optimization) & MW \\
$V_{gas}$ & Primary fueling gas valve voltage (zeroed for trajectory optimization) & Volts \\
\addlinespace
\midrule
\multicolumn{3}{c}{\textbf{Predicted Observations} $\mathbf{o}_t$} \\
\addlinespace
$W_{tot}$ & Plasma total stored thermal energy & kJ \\
$\bar{n}_{e,20}V$ & Line-averaged electron density times volume & $10^{20}$ \\
$f_{GW}$ & Greenwald fraction & - \\
$\beta_p$ & Plasma poloidal beta & - \\
$q_{95}$ & Safety factor at the 95\% flux surface & - \\
$\gamma_{vgr}$ & Vertical instability growth rate & 1000$s^{-1}$ \\
$T_e(\rho)$ & Electron temperature profile (not predict-first) & keV \\
$n_{e, 20}(\rho)$ & Electron density profile (not predict-first) & $10^{20} \, \text{m}^{-3}$ \\
\bottomrule
\end{tabularx}
\caption{The set of observations predicted by the learned dynamics model in response to action inputs. The time derivatives of certain quantities are chosen as the action to allow the ramp-rate to be constrained during trajectory optimization. Note that certain actions are not optimized in this work, but are important to include for the purpose of training the dynamics model; these actions are simply set to zero during the trajectory optimization process. Also note that the set of inputs used for the profile predictor component is $[W_{tot}, \bar{n}_{e,20}, P_{NBI}, P_{ECRH}, I_p, \kappa, a_{minor}, \delta]$.}
\label{tab:obs_action}
\end{table}

\newpage

\section*{Figures}

\begin{figure}[!htbp]
    \centering
    \includegraphics[width=\linewidth]{img/shot_breakdown.pdf}
    \caption{\textbf{A shot by shot breakdown of every shot in the experiment}. The results show the plasma current, $I_p$, and stored energy, $W_{tot}$, at time of plasma termination along with additional contextual information. The arbitrary shading is for distinguishing groups of shots described by the labels.}
    \label{fig:shot-breakdown}
\end{figure}

\begin{figure}[!htbp]
    \centering
    \includegraphics[width=0.8\linewidth]{img/170kA_predict.pdf}
    \caption{\textbf{A priori predictions and experimental results for the extrapolation test scenario}. \textbf{a}) Action trajectories and a priori predictions of plasma dynamics during rampdown, compared to experimental results from TCV \#82877 and \#82878. The RL training environment accounts for uncertainty in actuation with distributions on the action trajectory; the average of the distribution (in solid black) is used for shot programming. Control of the plasma shape proved to be a challenge for this phase, an issue also observed in previous rampdown studies \cite{mehta2024automated, van2023scenarioA}. Shown action variables include the plasma current, $I_p$, neutral beam injected power, $P_{NBI}$, minor radius, $a_{minor}$, elongation, $\kappa$, and high-field side gap, $g_{HFS}$. Shown predictions and constraints include the stored energy, $W_{tot}$, the line-averaged electron density, $\bar{n}_{e,20}$, poloidal beta, $\beta_p$, Greenwald fraction, $f_{GW}$, rotational transform at the 95\% flux surface, $\iota_{95}$, and vertical instability growth rate, $\gamma_{vgr}$. \textbf{b}) Post-hoc predictions of electron temperature, $T_e$, and density, $n_e$, profiles compared to Thomson Scattering measurements.}
    \label{fig:170_predict_first}
\end{figure}

\newpage


\setcounter{figure}{0}
\renewcommand{\thefigure}{S\arabic{figure}}
\renewcommand{\thetable}{S\arabic{table}}
\section{Supplementary Information}
\begin{table}[!htbp]
\centering
\begin{tabularx}{\textwidth}{@{}lXl@{}}
\toprule
\textbf{Random Variable} & \textbf{Distribution (140kA/170kA scenarios)} & \textbf{Quantification Method} \\
\midrule
\multicolumn{3}{c}{\textbf{Disturbances}} \\
\addlinespace
H-L back-transition threshold mult. factor & $N(\mu=1.15,\sigma=0.15)$  / $N(\mu=1.15,\sigma=0.1)$ & Ad-hoc\\
Ohmic heating mult. factor & $N(\mu=1, \sigma=0.1)$ / $N(\mu=1, \sigma=0.1)$ & Model prediction error\\
Radiated power mult. factor & $N(\mu=1, \sigma=0.5)$ / $N(\mu=1, \sigma=0.5)$ & Model prediction error \\
Energy confinement time mult. factor & $N(\mu=1, \sigma=0.075)$ / $N(\mu=1, \sigma=0.075)$ & Model prediction error \\
Particle confinement time mult. factor & $N(\mu=1, \sigma=0.075)$ /  $N(\mu=1, \sigma=0.075)$ & Model prediction error \\
Wall effects NN mult. factor & $N(\mu=1, \sigma=0.2)$ / $N(\mu=1, \sigma=0.2)$ & Ad-hoc \\
$\gamma_{vgr}$ mult. factor & $N(\mu=1,\sigma=0.2)$ / $N(\mu=1,\sigma=0.2)$& Model prediction error \\
$g_{HFS}$ actuation mult. factor  & $N(\mu=1, \sigma=0.5)$ / $N(\mu=1, \sigma=0.5)$ & Database Control Errors \\
Limited $\tau_E$ mult. factor & $U(1, 1.15)$ / $U(1, 1.2)$ & Ad-hoc \\
Limited $\tau_n$ mult. factor & $U(1, 1.15)$ / $U(1, 1.2)$  & Ad-hoc \\
H-mode correction factor for $\tau_n$ & $U(0.3, 0.5)$ / $U(0.3, 0.5)$  & Ad-hoc \\
Jump in $\kappa$ at the diverted to limited transition & $U(-0.2, 0.0)$ / $U(-0.2, 0.0)$ & Database Control Errors \\
\midrule
\multicolumn{3}{c}{\textbf{L-mode Initial State}} \\
\addlinespace
$I_p$ & $N(\mu=1.4, \sigma=0.025)$ / $N(\mu=1.7, \sigma=0.04)$  & Database Control Errors \\
$W_{tot}$ & $N(\mu=10, 0.5)$ / $N(\mu=12, 0.5)$ & Database Control Errors \\
$\bar{n}_{e, 20} V$ & $N(\mu=0.8,\sigma=0.05)$ / $N(\mu=0.8,\sigma=0.05)$ & Database Control Errors \\
$\kappa$ & $N(\mu=1.575, 0.03)$ / $N(\mu=1.575, 0.03)$ & Database Control Errors \\
$a_{minor}$ & $N(\mu=0.21, 0.005)$ / $N(\mu=0.21, 0.005)$ & Database Control Errors \\
$\delta$ & $N(\mu=0.275, 0.01)$ / $N(\mu=0.275, 0.01)$ & Database Control Errors \\
$P_{NBI}$ & $U(1.05, 1.15)$ / $U(1.1, 1.2)$ & Database Control Errors \\

\midrule
\multicolumn{3}{c}{\textbf{H-mode Initial State}} \\
\addlinespace
$I_p$ & $N(\mu=1.4, \sigma=0.025)$ / $N(\mu=1.7, \sigma=0.04)$& Database Control Errors \\
$W_{tot}$ & $N(\mu=12.5, 0.5)$ / $N(\mu=16, 0.5)$ & Database Control Errors\\
$\bar{n}_{e, 20} V$ & $N(\mu=0.9,\sigma=0.05)$ / $N(\mu=1.1,\sigma=0.05)$ & Database Control Errors\\
$\kappa$ & $N(\mu=1.575, 0.03)$ / $N(\mu=1.575, 0.03)$ & Database Control Errors\\
$a_{minor}$ & $N(\mu=0.21, 0.005)$ / $N(\mu=0.21, 0.005)$ & Database Control Errors \\
$\delta$ & $N(\mu=0.275, 0.01)$ / $N(\mu=0.275, 0.01)$ & Database Control Errors \\
$P_{NBI}$ & $U(1.05, 1.15)$ / $U(1.1, 1.2)$ & Database Control Errors \\
\addlinespace
\bottomrule
\end{tabularx}
\caption{\textbf{RL uncertainty model}. The uncertainty model used in RL training environments to design trajectories with robustness to distributional uncertainty for both the 140kA and 170kA scenarios.}
\label{tab:uncertainty_model}
\end{table}
\begin{table}[!htbp]
  \centering
  \begin{minipage}[t]{0.48\textwidth} 
    \centering
    \begin{tabular}{llr}
        \toprule
        \textbf{Category} & \textbf{Parameter} & \textbf{Value} \\
        \midrule
        \multirow{5}{*}{Optimizer} 
          & Initial Learning Rate & 0.02 \\
         & Final Learning Rate & 0.0001 \\
         & Transition Steps & 200\\
         & Decay Rate & 0.9\\
         & Weight Decay & 0.005\\
        \midrule
        \multirow{2}{*}{$NN_{ohm,rad}$} 
          & Depth & 2\\
          & Width & 32\\
        \midrule
      \multirow{2}{*}{$NN_{vgr}$} 
          & Depth & 3\\
          & Width & 256\\
       \midrule
      \multirow{2}{*}{$NN_{conf}$} 
          & Depth & 1\\
          & Width & 128\\
      \midrule
      \multirow{2}{*}{$NN_{prof}$} 
          & Depth & 2\\
          & Width & 64\\
      \bottomrule
    \end{tabular}
  \end{minipage}
  \hfill 
  \begin{minipage}[t]{0.48\textwidth} 
    \centering
    \begin{tabular}{llr}
        \toprule
        \textbf{Category} & \textbf{Parameter} & \textbf{Value} \\
        \midrule
        \multirow{2}{*}{Hard Limits} & $f_{GW}$ & 1.0 \\
         & $\iota_{95}$ & 0.5 \\
        \midrule
        \multirow{4}{*}{Soft Limits} & $f_{GW}$ & 0.8 \\
         & $\beta_p$ & 1.75 \\
         & $\gamma_{vgr}$ & 0.75 \\
         & $\iota_{95}$ & 0.313 \\
        \midrule
        \multirow{5}{*}{Parameters} 
         & $c_{time}$ & 5.0 \\
         & $c_{I_p}$ & 1.0 \\
         & $c_W$ & 1.0 \\
         & $c_{soft}$ & 1.0×10\textsuperscript{3} \\
         & $c_{hard}$ & 5.0×10\textsuperscript{4} \\
        \bottomrule
    \end{tabular}
  \end{minipage}
  \captionof{table}{\textbf{Hyperparameters and reward function parameters.} (Left) Hyperparameters for the final trained model. (Right) Reward function parameters. Note that the values corresponding to limits are the maximum allowed values.} 
  \label{tab:hyperparams_reward_params}
\end{table}

\newpage
\begin{figure}[!htbp]
    \centering
    \includegraphics[width=0.9\linewidth]{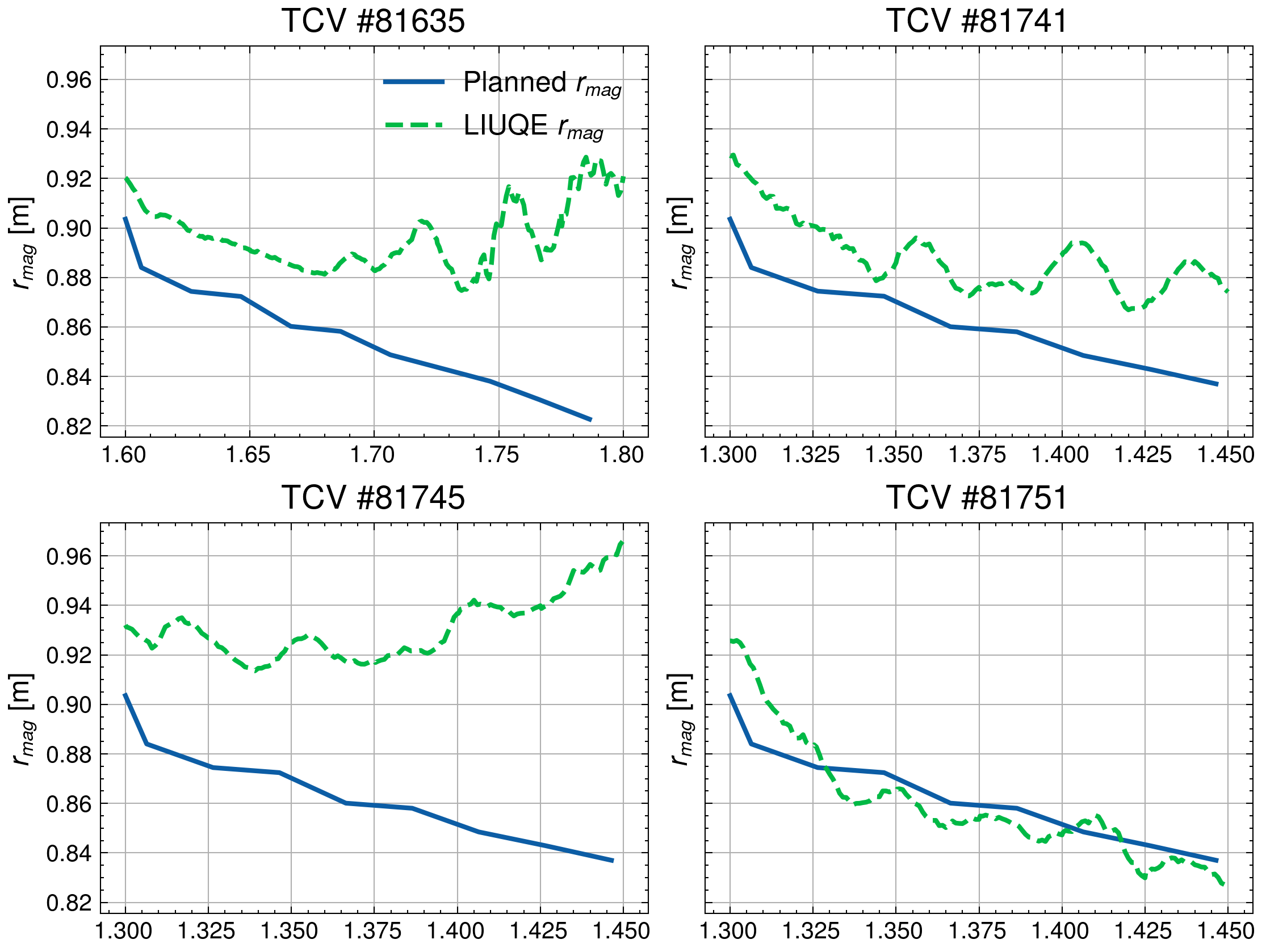}
    \caption{\textbf{Debugging the legacy radial observer issue}. Plot of planned radial magnetic position $r_{mag}$ against observed values from equilibrium reconstruction with LIUQE, showing exceptionally poor control performance during initial rampdown attempts. The issue was identified as a legacy $\frac{1}{I_p^2}$ term used in generating the radial position observer feed-forward trajectory, with the issue finally fixed in TCV\#81751.}
    \label{fig:rmag-debug}
\end{figure}

\newpage
\begin{figure}
    \centering
    \includegraphics[width=0.9\linewidth]{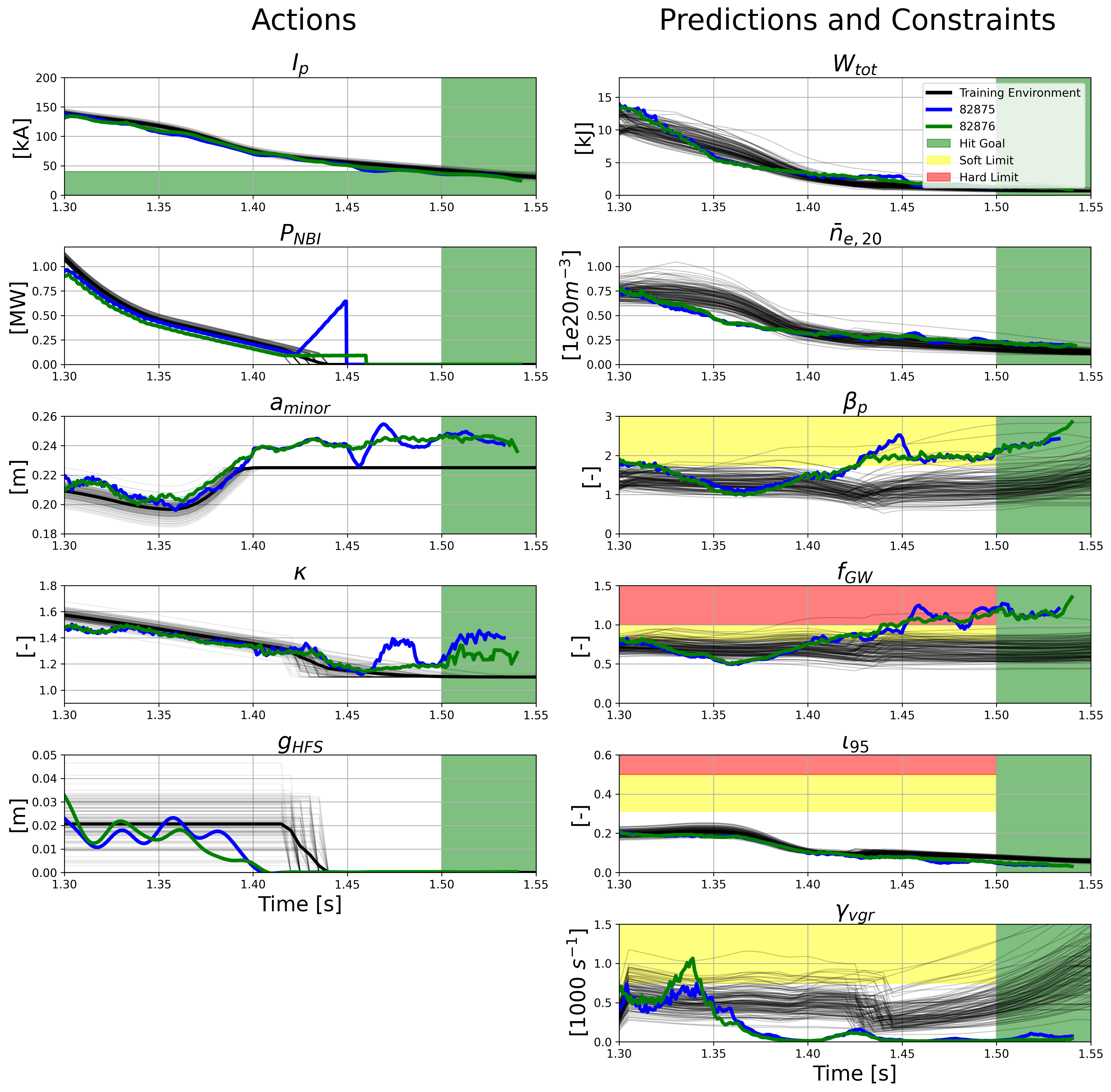}
    \caption{\textbf{Results from the 140kA Scenario}. Shown are a priori predictions of the RL environment and experimental outcomes of the final two 140kA shots.}
    \label{fig:140ka_pred}
\end{figure}

\newpage
\begin{figure}
    \centering
    \includegraphics[width=0.75\linewidth]{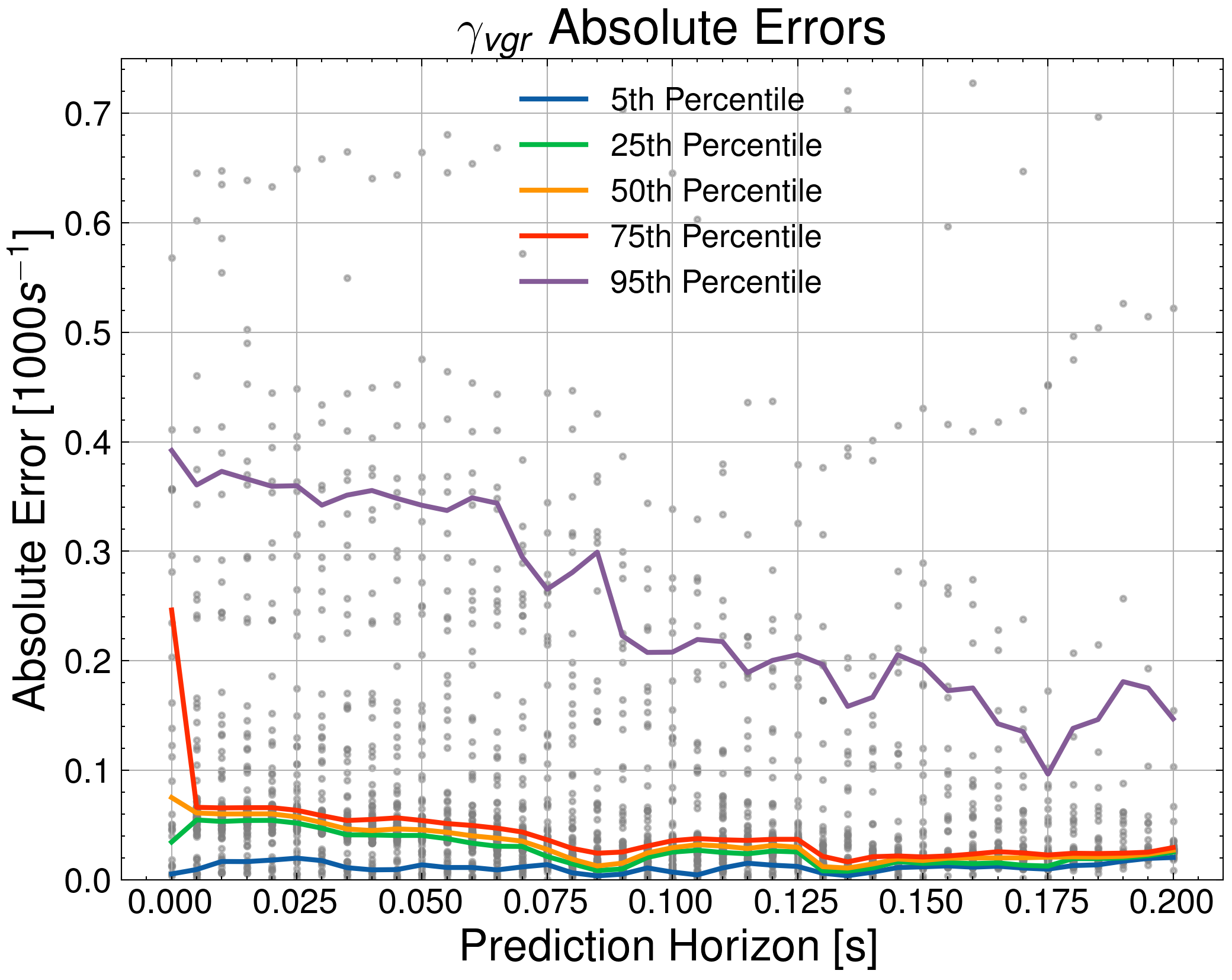}
    \caption{\textbf{Absolute errors for $\gamma_{vgr}$}. Errors are shown for the initial validation set of 131 shots, showing relatively low absolute prediction error in the majority of shots.}
    \label{fig:vgr-abs-err}
\end{figure}
\newpage
\begin{figure}[!htbp]
    \centering
    \includegraphics[width=0.85\textwidth]{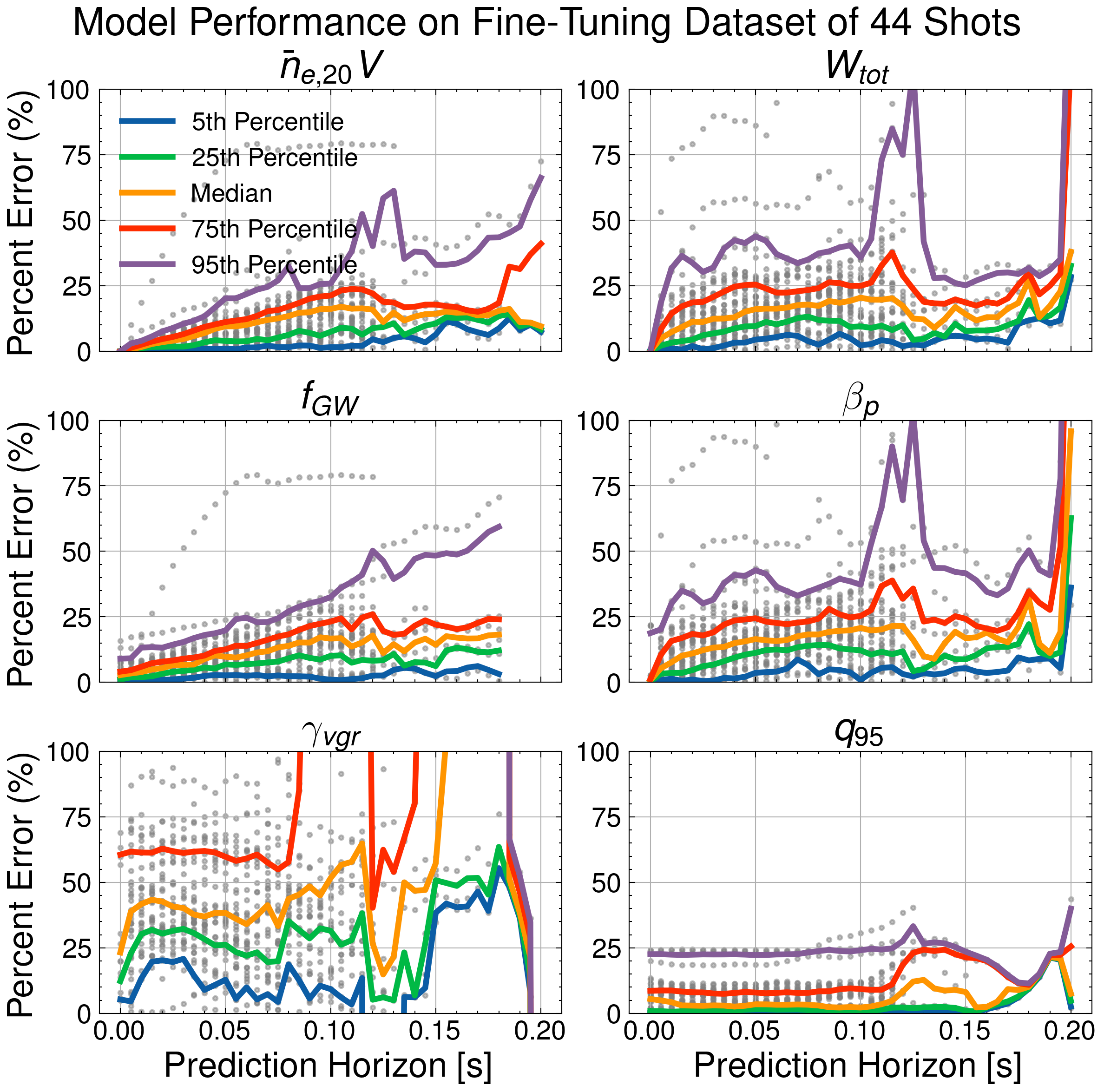}
    \caption{\textbf{Fine-tuning dataset metrics}. Model prediction accuracy as a function of prediction horizon during rampdowns on the fine-tuning dataset of 44 shots.}
    \label{fig:fine_tune_val}
\end{figure}
\newpage
\begin{figure}[!htbp]
    \centering
\includegraphics[width=0.75\linewidth]{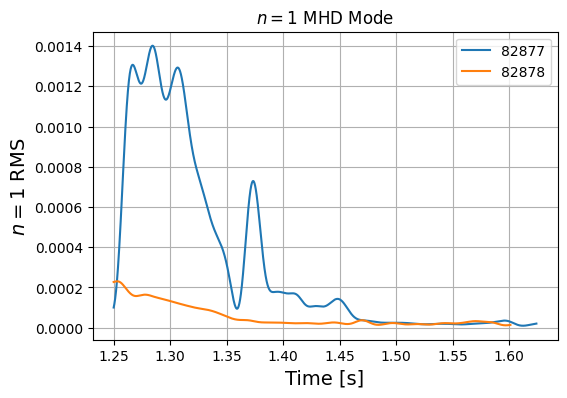}
    \caption{\textbf{NTM driven discrepancy between \#82877 and \#82878}. Shown is the $n=1$ root mean squared (RMS) signal derived from magnetic diagnostics indicating the presence of a neo-classical tearing mode at the beginning of ramp-down for \#82877 that is not present in \#82878.}
    \label{fig:ntm}
\end{figure}
\newpage
\begin{figure}[!htbp]
    \centering
    \includegraphics[width=\linewidth]{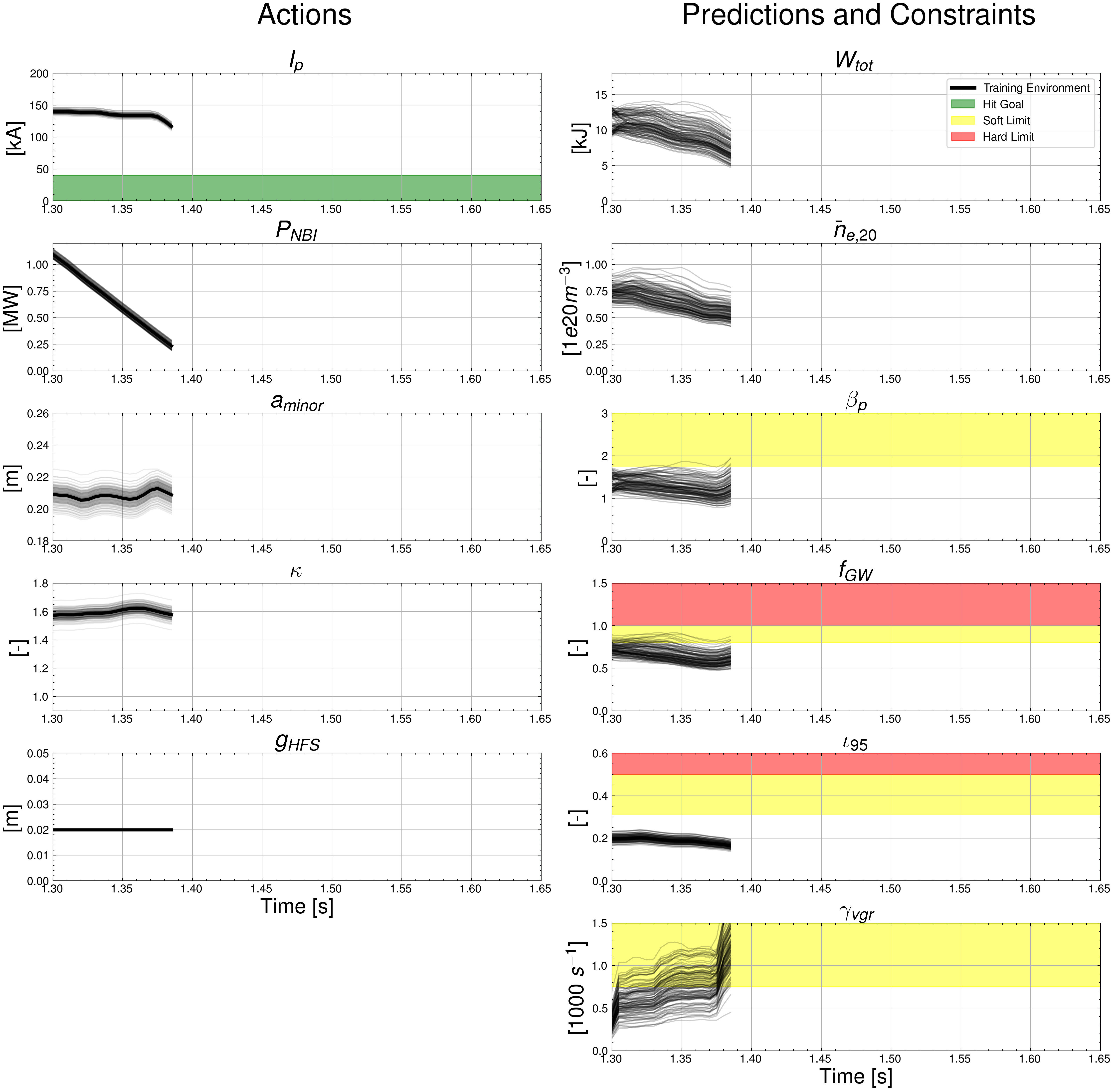}
    \caption{\textbf{\#81101 counterfactual analysis}. A counterfactual analysis, using the action trajectories from the disruptive rampdown, \#81101, in the RL training environment used for trajectory design. We see that the training environment predicts a large $\gamma_{vgr}$, well within the soft limit region right before the shot disrupted in experiment.}
    \label{fig:act81101_counterfactual}
\end{figure}

\begin{figure}
    \centering
    \includegraphics[width=\linewidth]{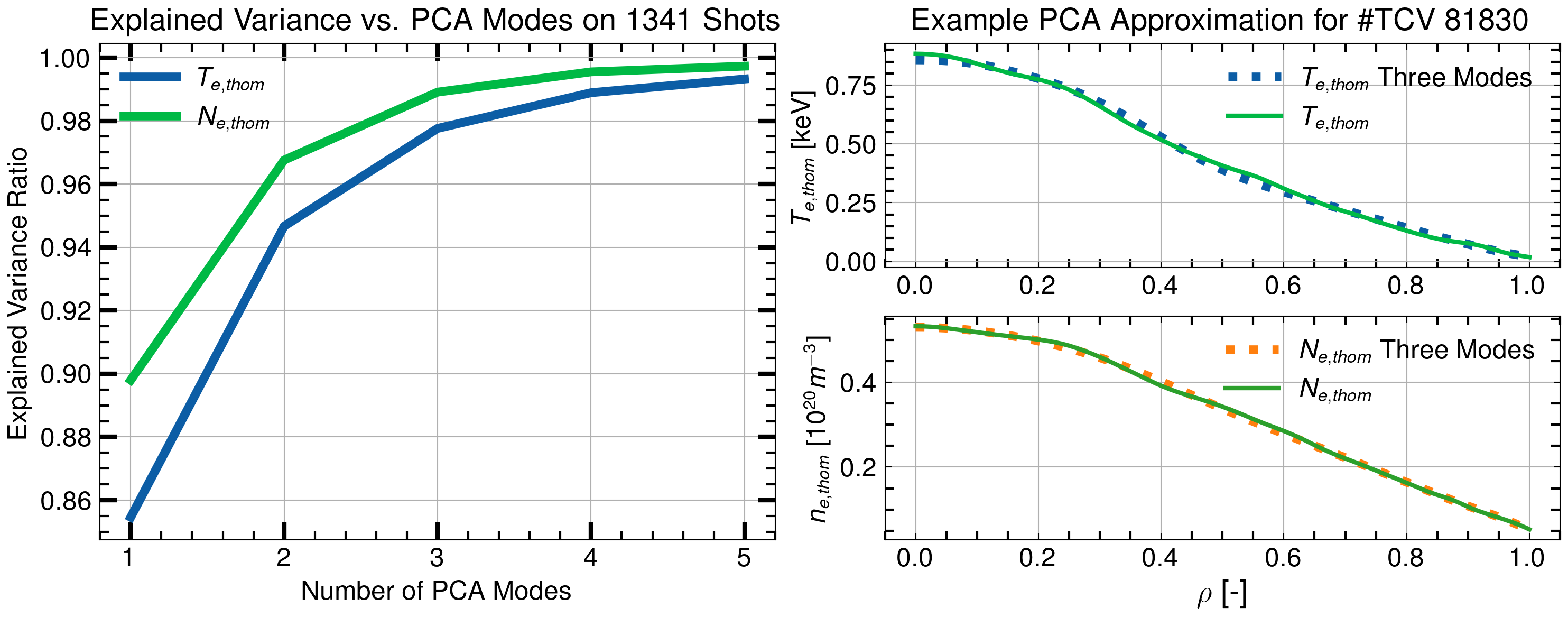}
    \caption{\textbf{Principal component analysis (PCA) showing the empirical low dimensionality of kinetic profiles measured by Thomson Scattering}. (Left) explained variance as a function of number of PCA modes, showing high explained variance with a relatively low dimensional representation of kinetic profiles on a database of 1341 shots. (Right) an example of kinetic profiles and their corresponding approximation with three PCA modes.}
    \label{fig:pca-ev-profiles}
\end{figure}

\begin{figure}[!htbp]
    \centering
    \includegraphics[width=0.9\linewidth]{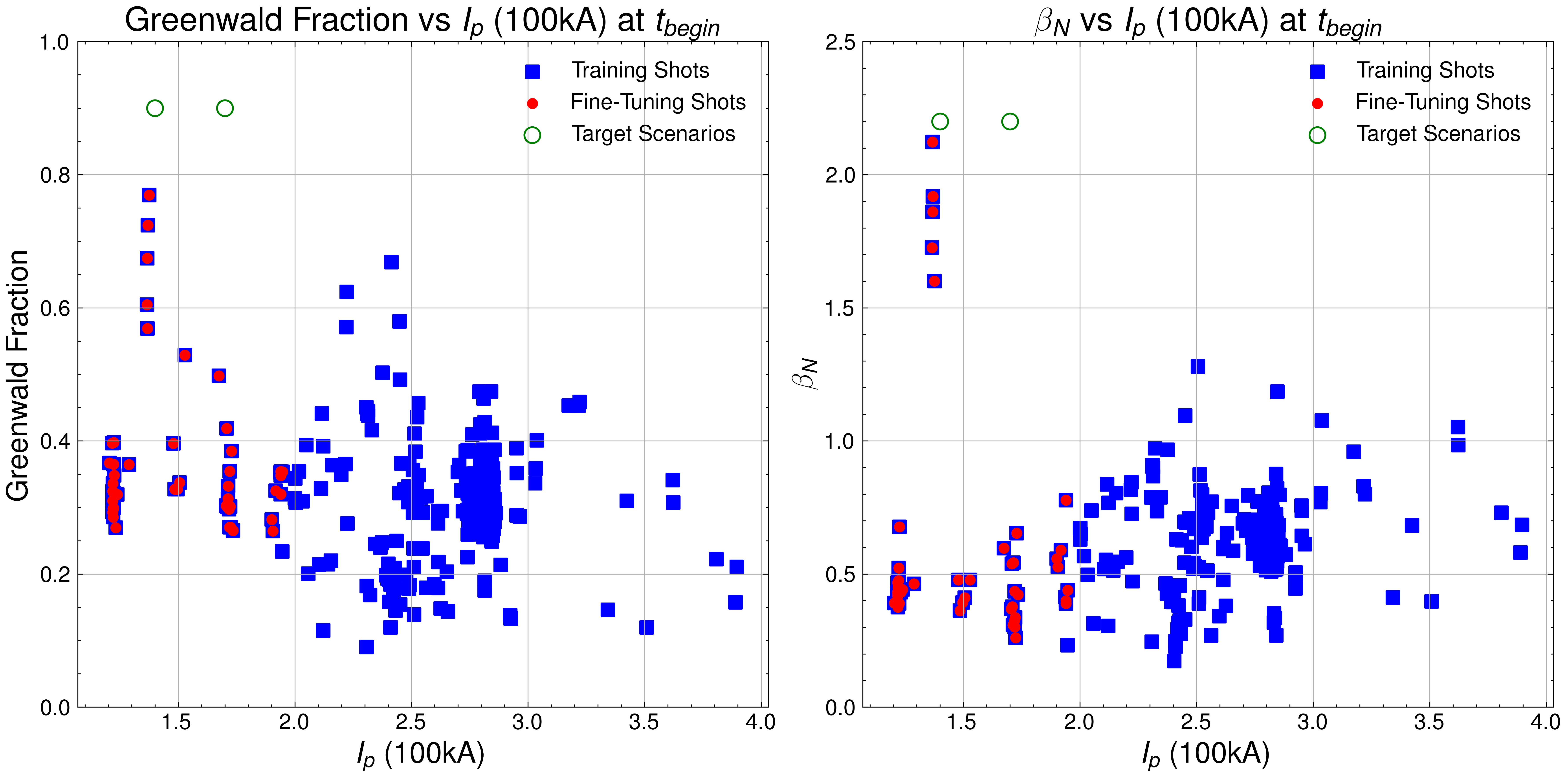}
    \caption{\textbf{The data distribution used for training the NSSM}. Each point corresponds to a shot, visualized in ($I_p$, $f_{GW}$) space and ($I_p$, $\beta_N$) space.}
    \label{fig:data-distribution}
\end{figure}
\newpage
\begin{figure}[!htbp]
    \centering
    \includegraphics[width=\linewidth]{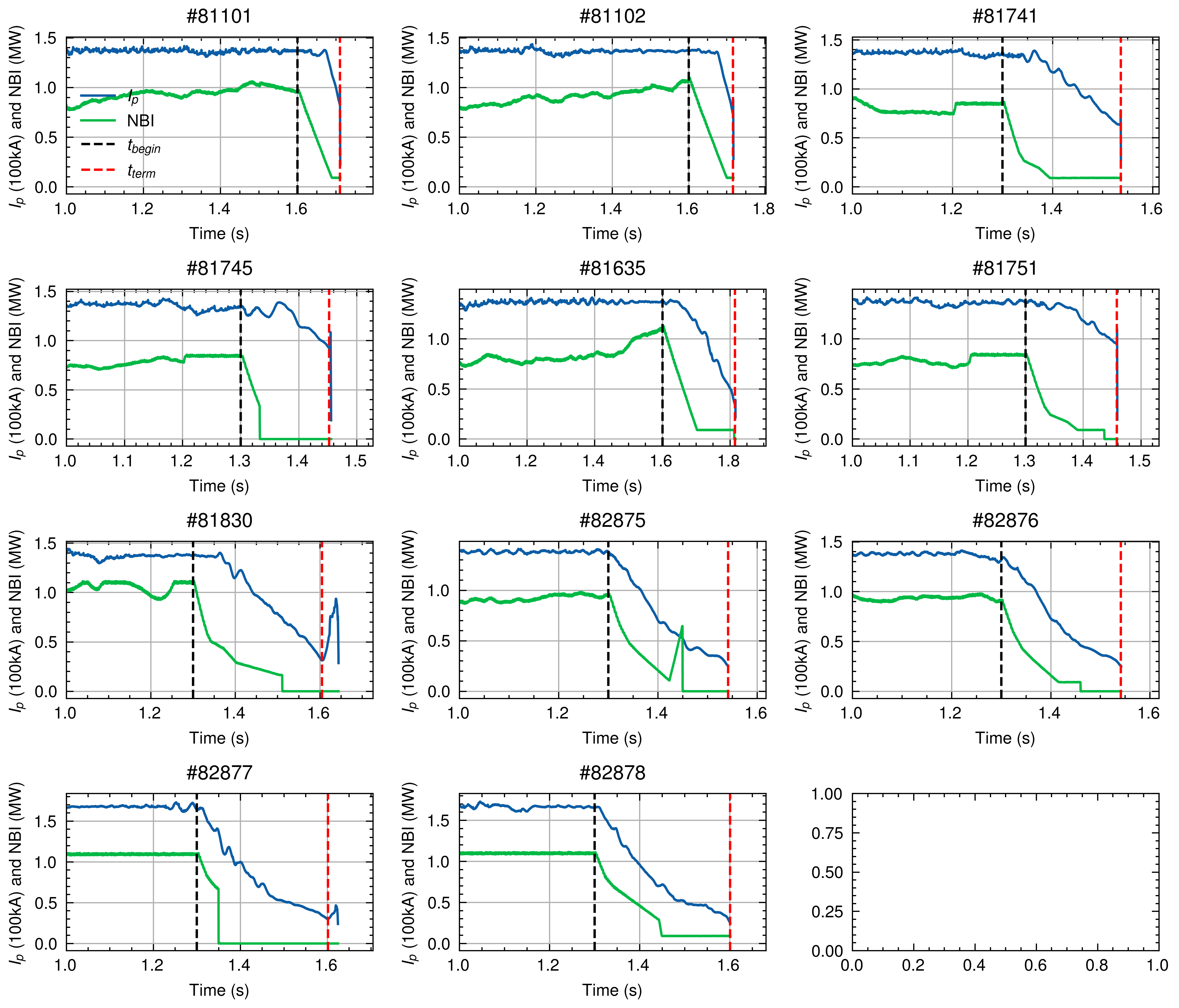}
    \caption{\textbf{Definition of the beginning and end of the termination phase for every shot in this experiment.} Note that in 81830, a low level software issue triggered an unexpected spike in plasma current at the very end of the rampdown. We define the termination time for that shot as the lowest achieved plasma current, as it is expected that net energy tokamaks where disruptions are of concern will have protection systems that prevent such incidents.}
    \label{fig:def_begin_end}
\end{figure}

\end{document}